\documentclass[twocolumn,aps,showpacs,superscriptaddress]{revtex4-1}
\usepackage{amsmath}
\usepackage{graphicx}
\RequirePackage[pagewise,mathlines]{lineno}

\begin{document}
\title{Double-slit experiment at fermi scale: coherent photoproduction in heavy-ion collisions}%
\author{W. Zha}\email{wangmei@rcf.rhic.bnl.gov}\affiliation{University of Science and Technology of China, Hefei, China}
\author{L. Ruan}\affiliation{Brookhaven National Laboratory, New York, USA}
\author{Z. Tang}\email{zbtang@ustc.edu.cn}\affiliation{University of Science and Technology of China, Hefei, China}
\author{Z. Xu}\affiliation{Brookhaven National Laboratory, New York, USA}\affiliation{Shandong University, Jinan, China}
\author{S. Yang}\affiliation{Brookhaven National Laboratory, New York, USA}
\date{\today}%
\begin{abstract}
The double-slit experiment has become a classic thought experiment, for its clarity in expressing the central puzzle of quantum mechanics -- wave-particle complementarity. Such wave-particle duality continues to be challenged and investigated in a broad range of entities with electrons, neutrons, helium atoms, C$_{60}$ fullerenes, Bose-Einstein condensates and biological molecules. All existing experiments are performed with the scale larger than angstrom. In this article, we present a double-slit scenario at fermi scale with new entities -- coherent photon products in heavy-ion collisions. Virtual photons from the electromagnetic fields of relativistic heavy ions can fluctuate to quark-antiquark pairs, scatter off a target nucleus and emerge as vector mesons. The two colliding nuclei can take turns to act as targets, forming a double-slit interference pattern. Furthermore, the ``which-way'' information can be partially solved by the violent strong interactions in the proposed scenario, which demonstrates a key concept of quantum mechanics -- complementary principle.
\end{abstract}
\maketitle
Relativistic heavy ions carry giant electromagnetic field which can be equivalent into a field of quasi-real virtual photons~\cite{KRAUSS1997503}. When two ions collide, the photon from the field of one nucleus may fluctuate into a virtual quark-antiquark pair which scatters elastically from the other nucleus, emerging as a real vector meson. The elastic scattering occurs via short-range strong force, which imposes a restriction on the production site of vector meson within one of the two ions. This provides the production process with a choice: either nucleus 1 emits a virtual photon while nucleus 2 acts as target, or vice versa. If the information is missing about which nucleus emits a photon or acts as target, the vector meson, behaving as a wave, originates simultaneously from both colliding nuclei. The wave-like behaviour and corresponding interference presents itself as a perfect double-slit experiment of individual particle. The double-slit experiment has become a classic thought experiment, for its clarity in expressing the central puzzle of quantum mechanics -- wave-particle complementarity. Such wave-particle duality~\cite{Broglie} continues to be challenged~\cite{Review_Scully,Review_Wiseman,PhysRevLett.95.040401,PhysRevLett.96.100403,nature_photonic} and investigated in a broad range of entities with electrons~\cite{Jnsson1961}, neutrons~\cite{RevModPhys.60.1067}, helium atoms~\cite{PhysRevLett.66.2689}, C$_{60}$ fullerenes~\cite{Review_C60}, Bose-Einstein condensates~\cite{Andrews637} and biological molecules~\cite{PhysRevLett.91.090408}. Currently, all existing experiments are performed with the scale larger than angstrom. In this article, we demonstrate that the coherent photoproduction in heavy-ion collisions could extend the double-slit scenario to fermi scale with new entities -- vector mesons~\cite{UPCreview}. Furthermore, the violent strong interactions accompanied with heavy-ion collisions could partially solve the ``which-slits'' problem, leading to the appearance of decoherence, which demonstrates a key concept of quantum mechanics -- complementary principle.

In contrast to the light vector mesons, photoproduction of heavy vector mesons offers an opportunity to directly determine the gluon distributions in nucleons and nuclei~\cite{Guzey:2013qza}, which are not directly accessible in deep inelastic scattering. Due to their clarity in physical picture and experimental feasibility, we choose J$/\psi$, the most abundant heavy vector meson, as the entity to illustrate the double-slit scenario in heavy-ion collisions. The J$/\psi$ photoproduction from the two nuclei is related by a parity transformation, and the parity of J$/\psi$ is negative, which assigns the amplitudes from the two directions with opposite signs. This gives the presented double-slit scenario a unique feature -- a phase shift of $\pi$ between the two interference sources. Furthermore the probability of multiple J$/\psi$ production in one collision is negligible in comparison with that of single production, which indicates that the interference pattern is build up one by one in the proposed scenario. This is an important version of double-slit experiments to clearly demonstrate the wave-particle duality, which states that all matter exhibits both wave and particle properties.
\renewcommand{\floatpagefraction}{0.75}
\begin{figure}[htbp]
\includegraphics[keepaspectratio,width=0.225\textwidth]{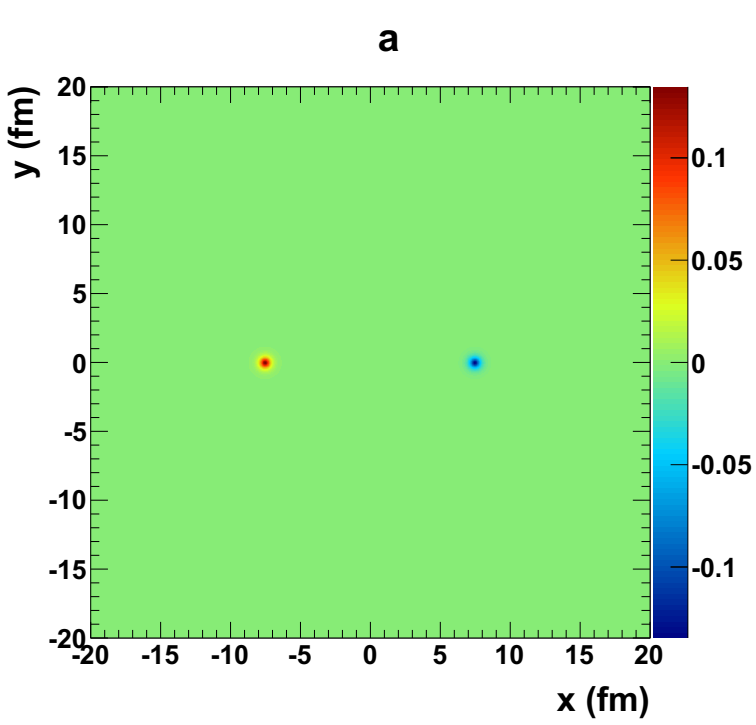}
\includegraphics[keepaspectratio,width=0.225\textwidth]{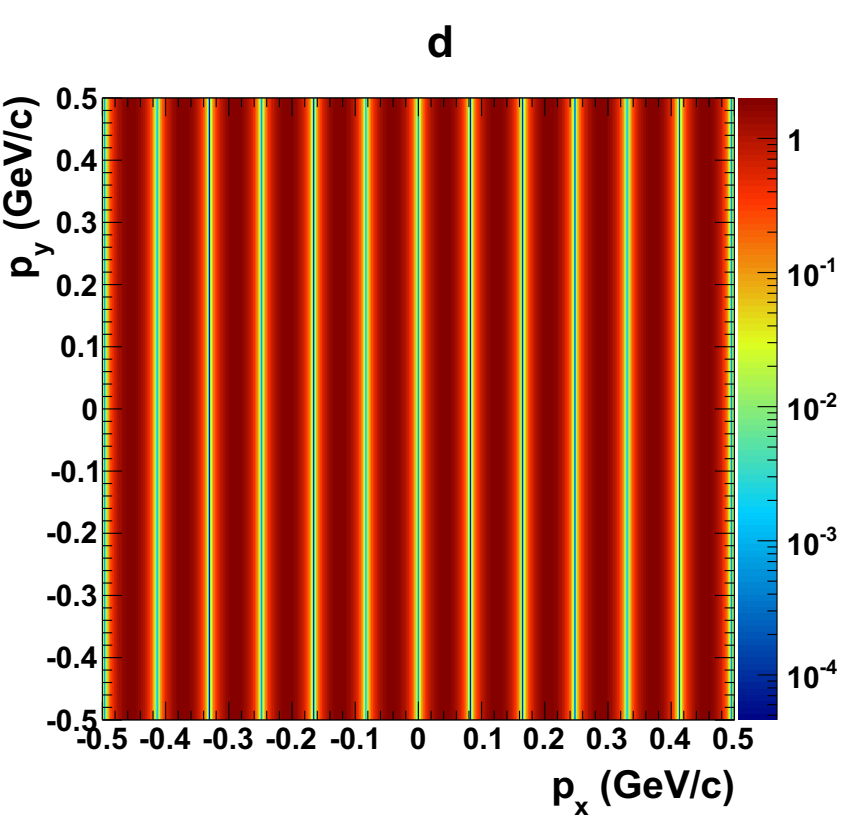}
\includegraphics[keepaspectratio,width=0.225\textwidth]{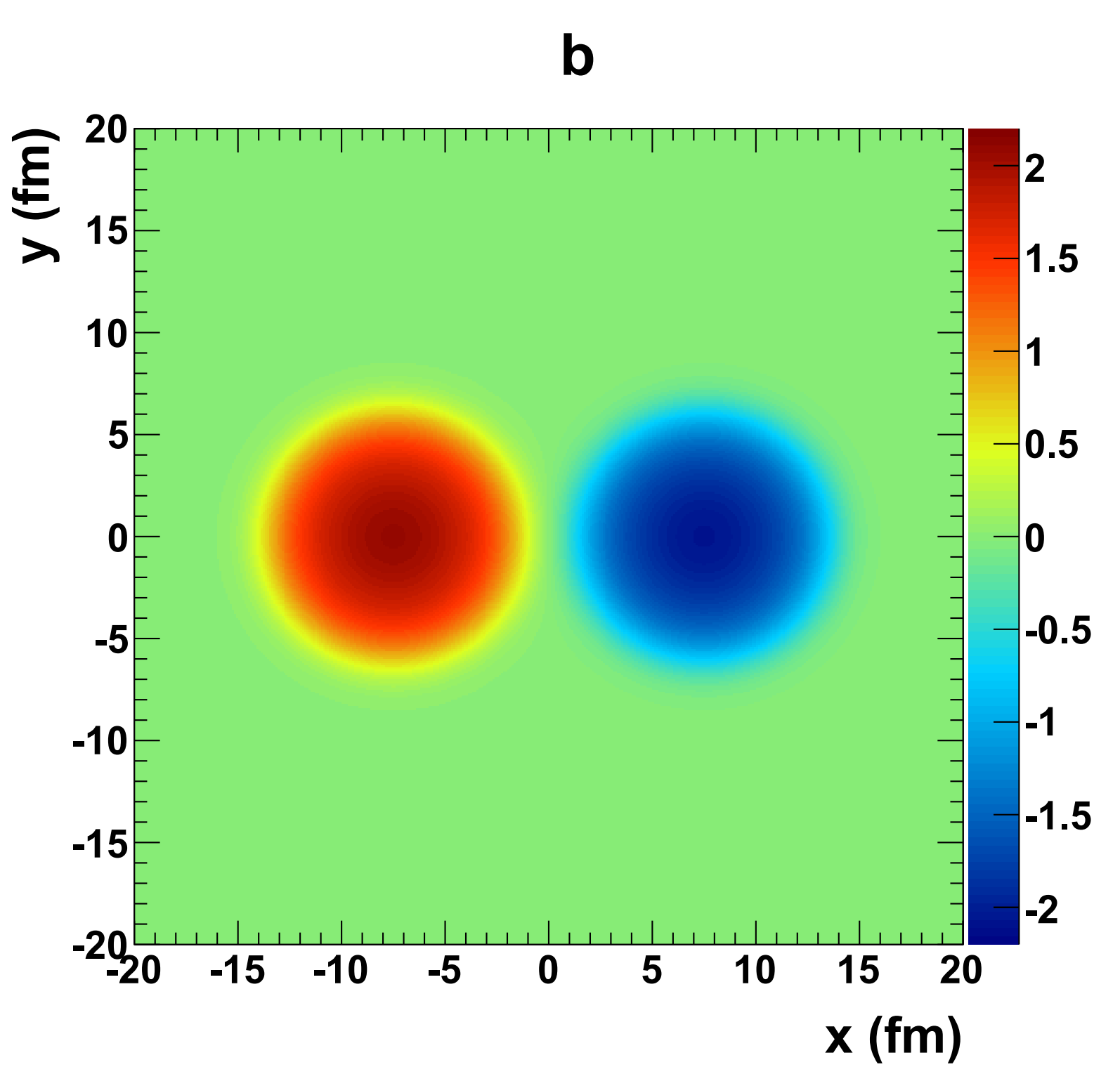}
\includegraphics[keepaspectratio,width=0.225\textwidth]{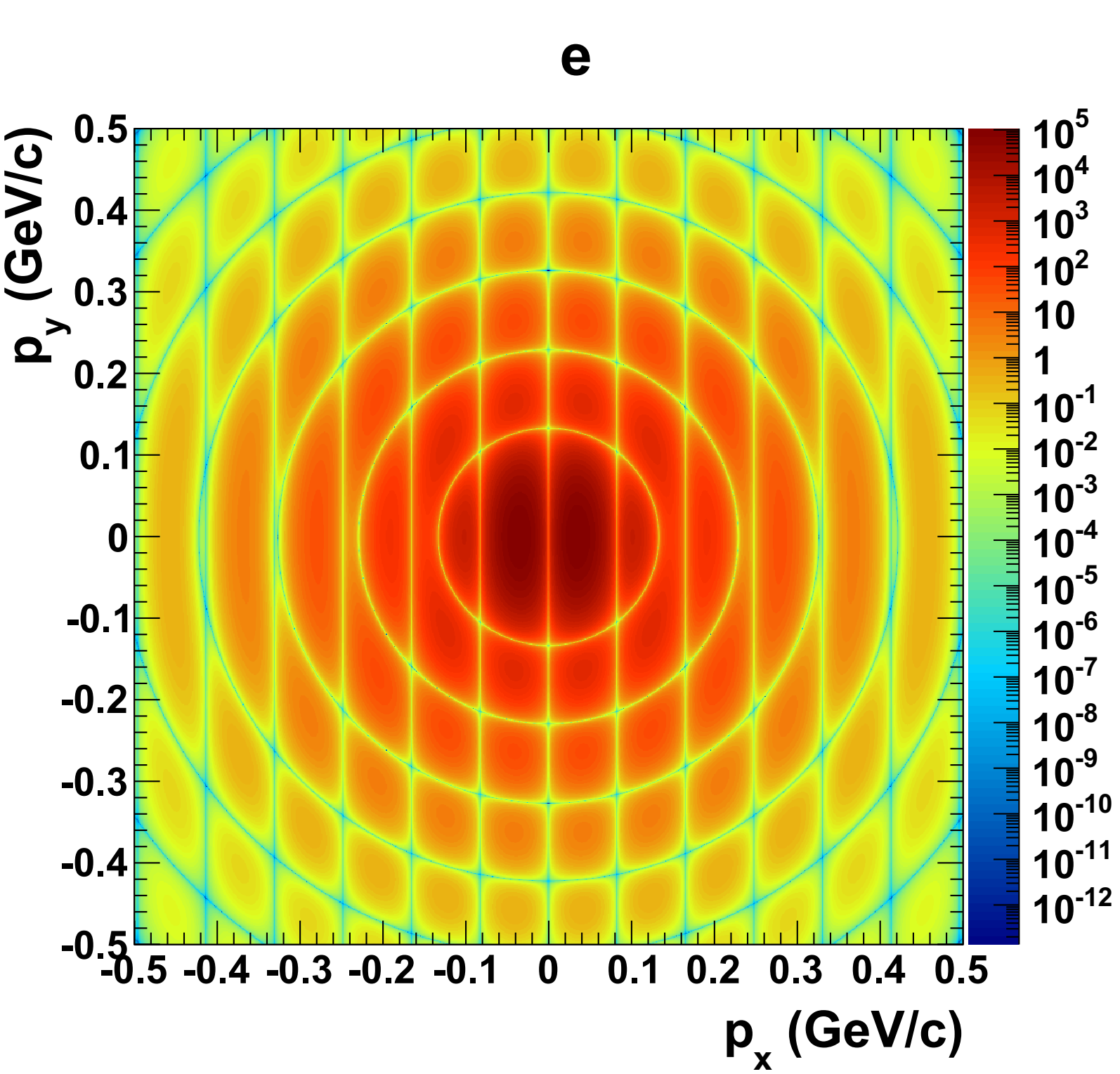}
\includegraphics[keepaspectratio,width=0.225\textwidth]{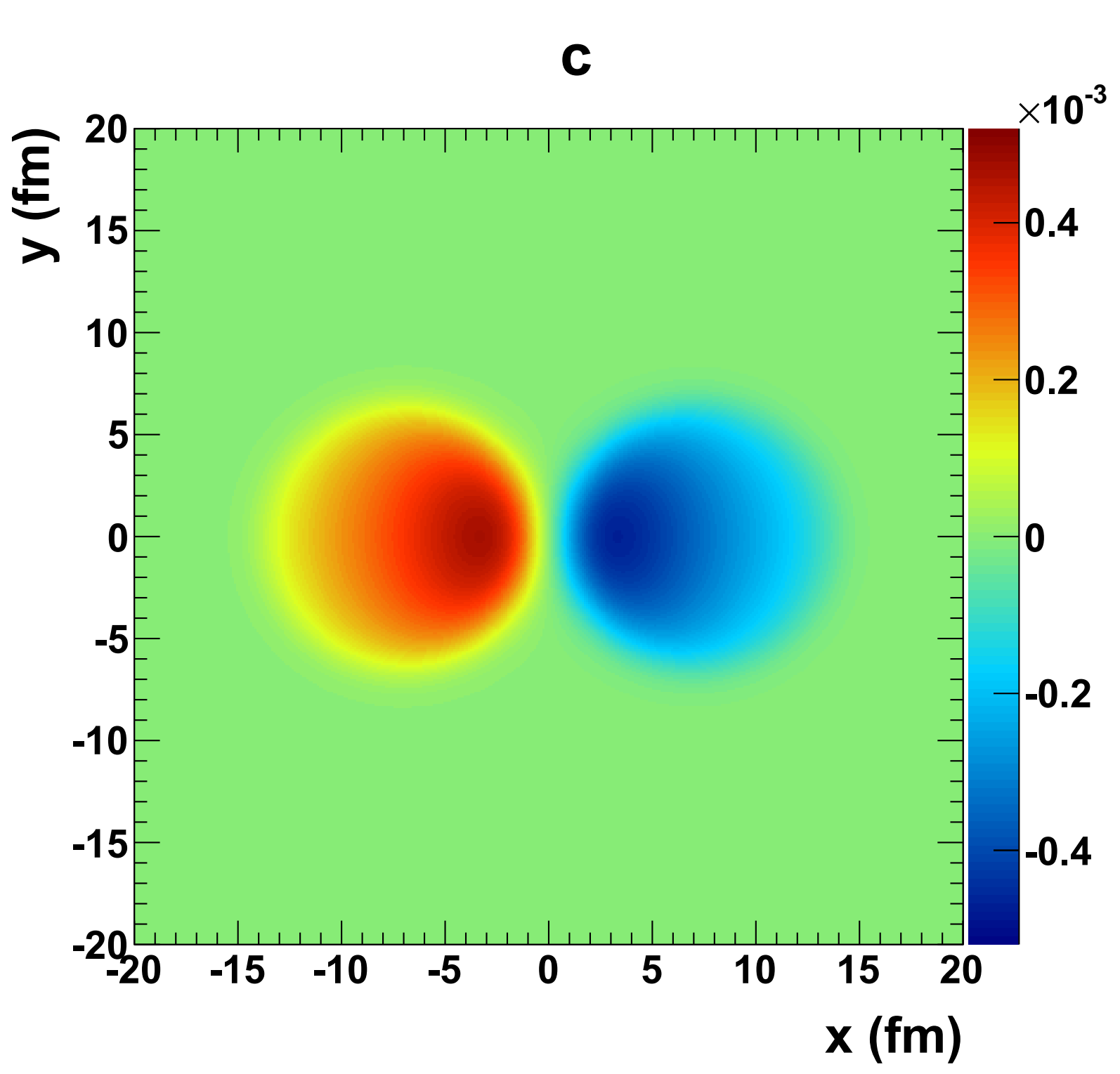}
\includegraphics[keepaspectratio,width=0.225\textwidth]{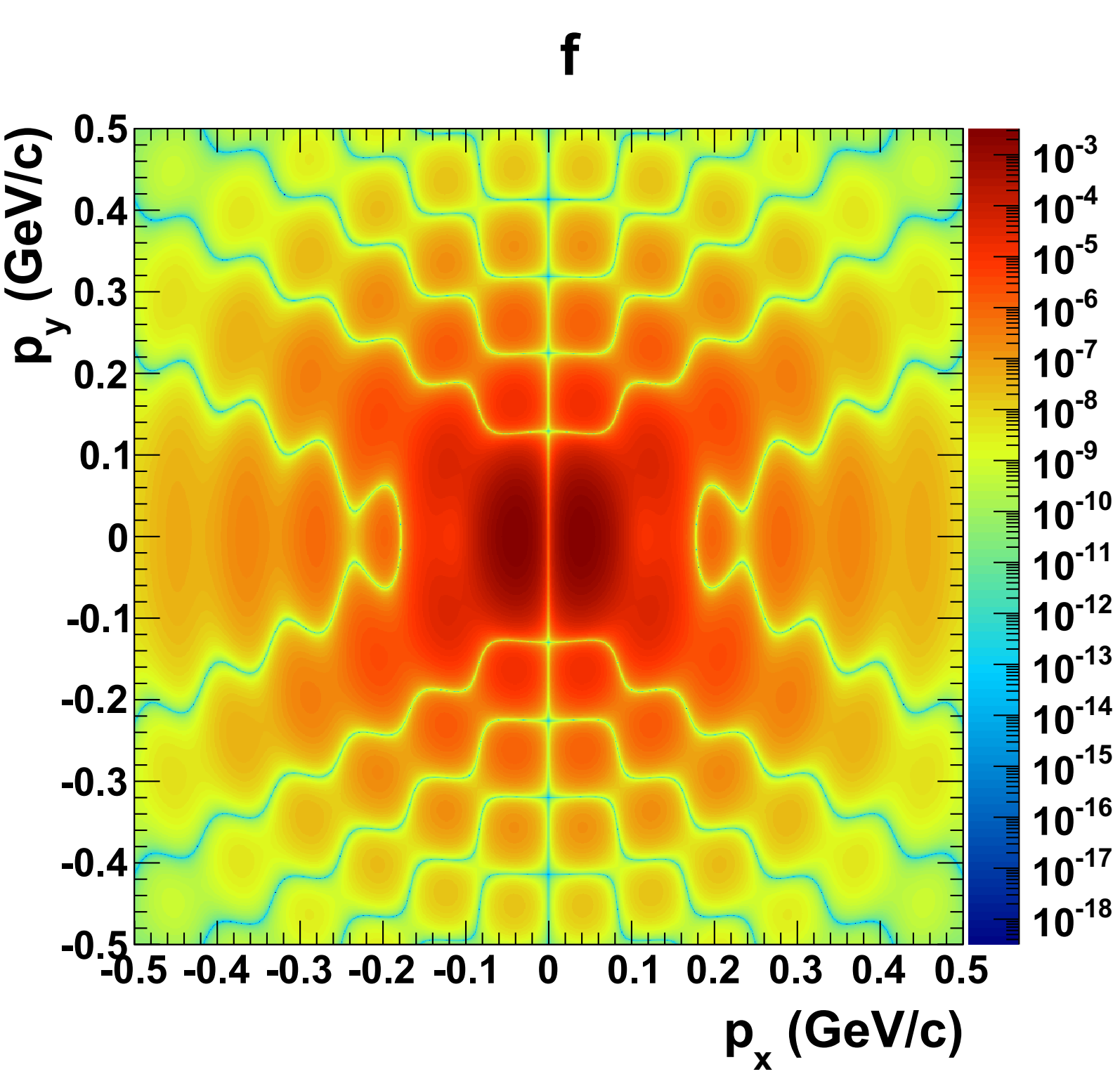}
\caption{Amplitude and momentum distribution patterns of coherent J/$\psi$ photoproduction with different scenarios for b = 15 fm at mid-rapidity (y=0). Panel a: point source scenario, panel b: Woods-Saxion distribution, panel c: realistic scenario described in the text. Panel d - f show the corresponding momentum distributions according to the amplitude scenarios in Panel a - c, respectively.}
\label{figure1}
\end{figure}
\begin{figure}[htbp]
\includegraphics[keepaspectratio,width=0.225\textwidth]{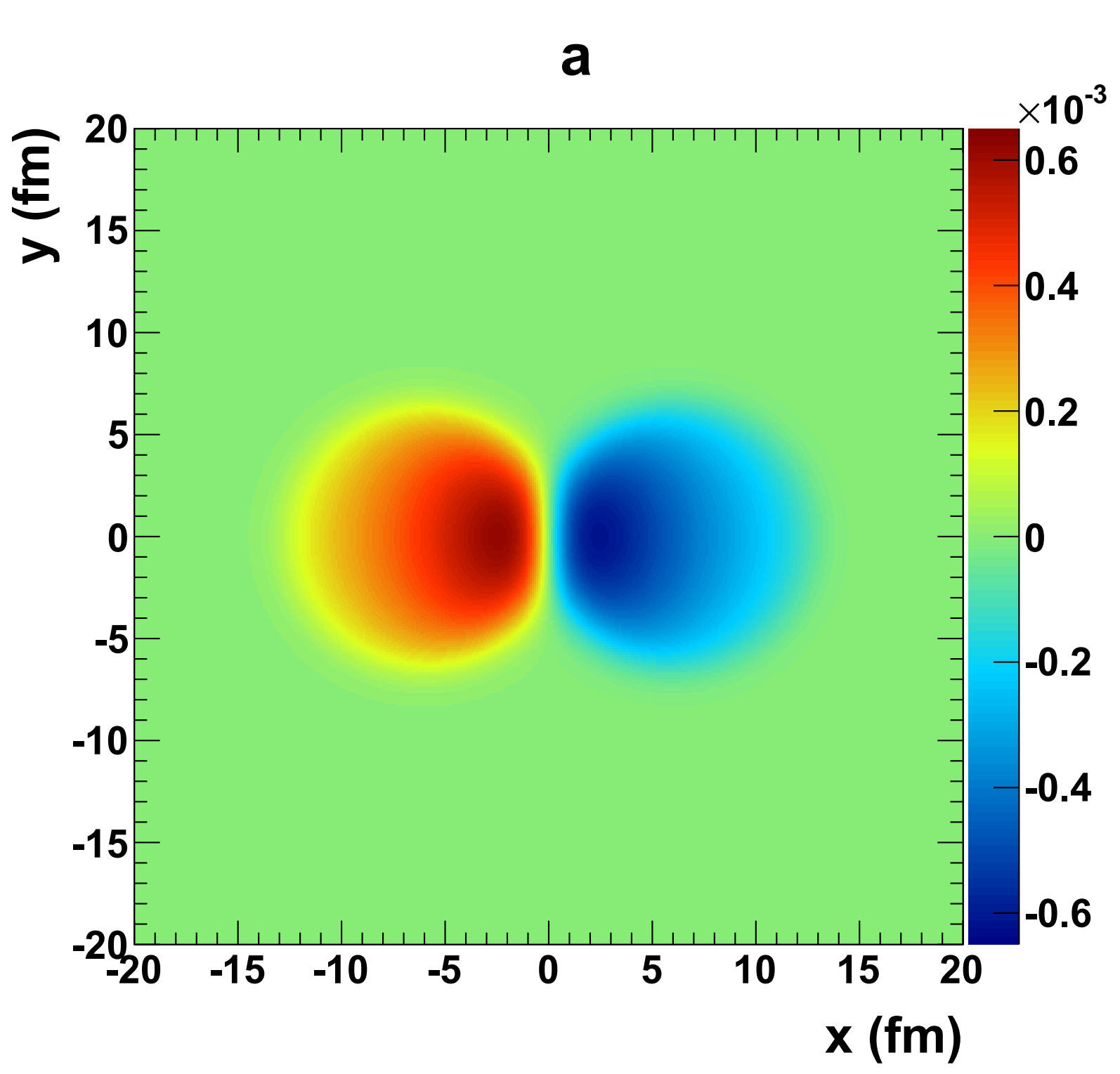}
\includegraphics[keepaspectratio,width=0.225\textwidth]{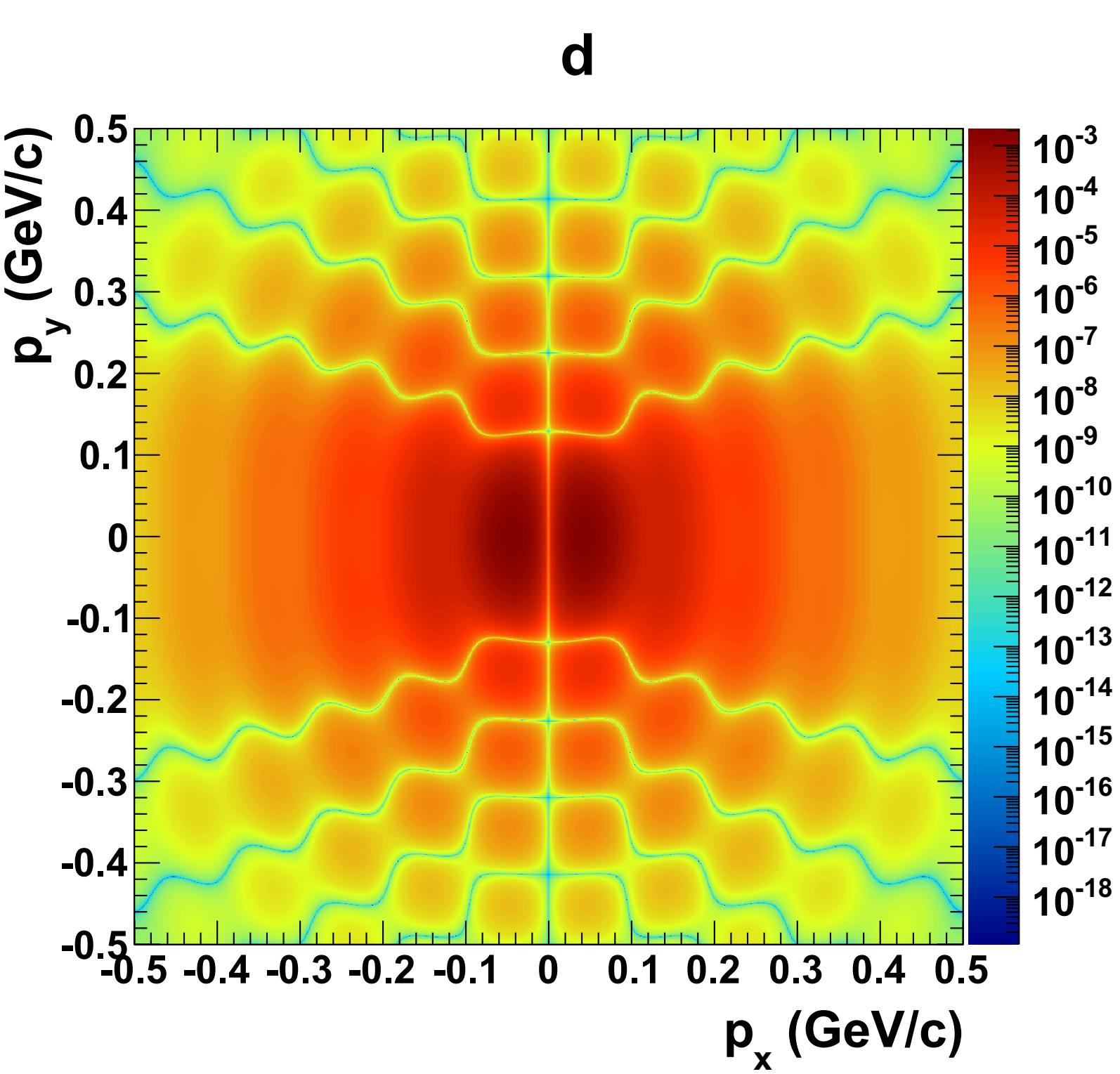}
\includegraphics[keepaspectratio,width=0.225\textwidth]{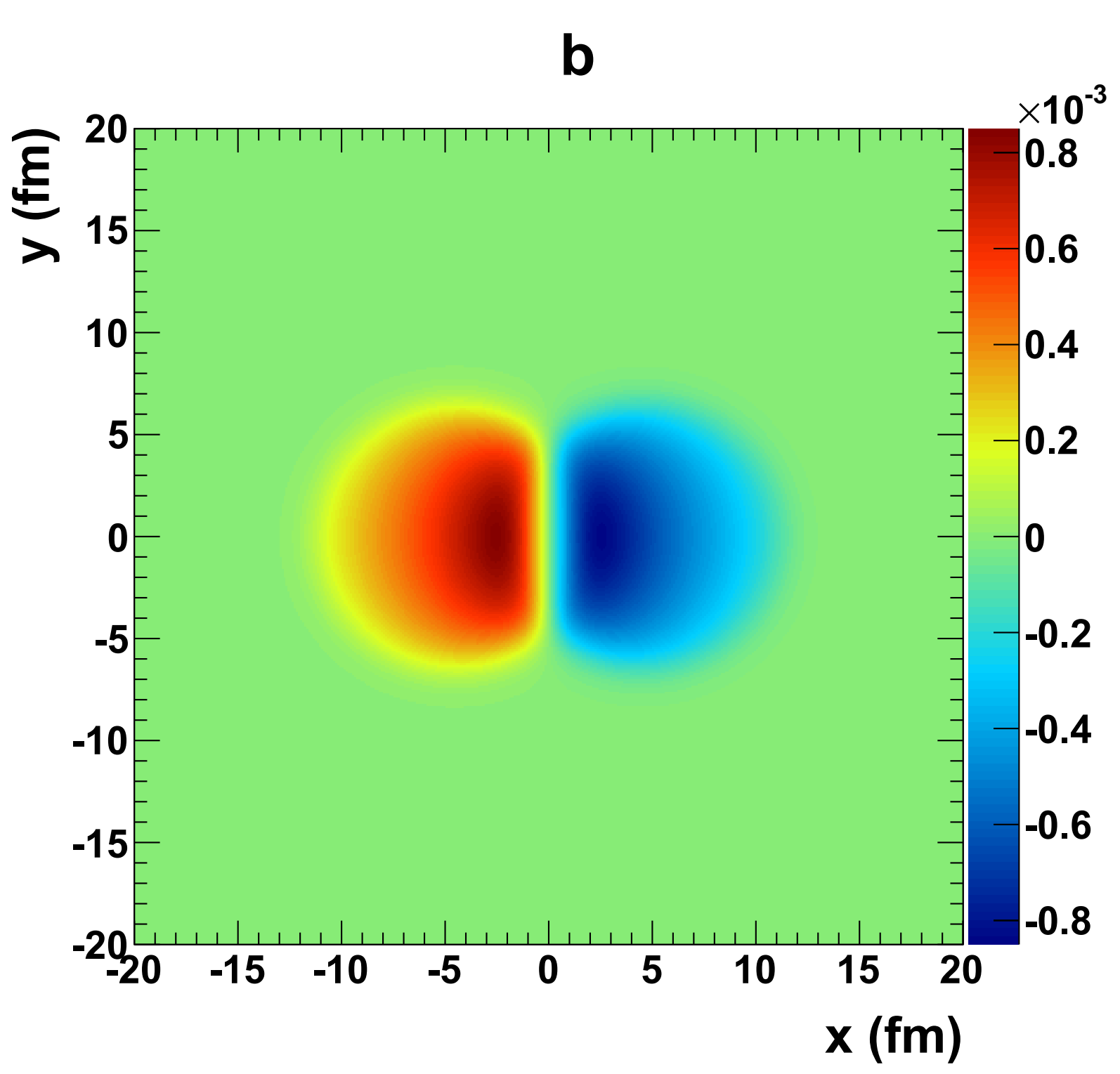}
\includegraphics[keepaspectratio,width=0.225\textwidth]{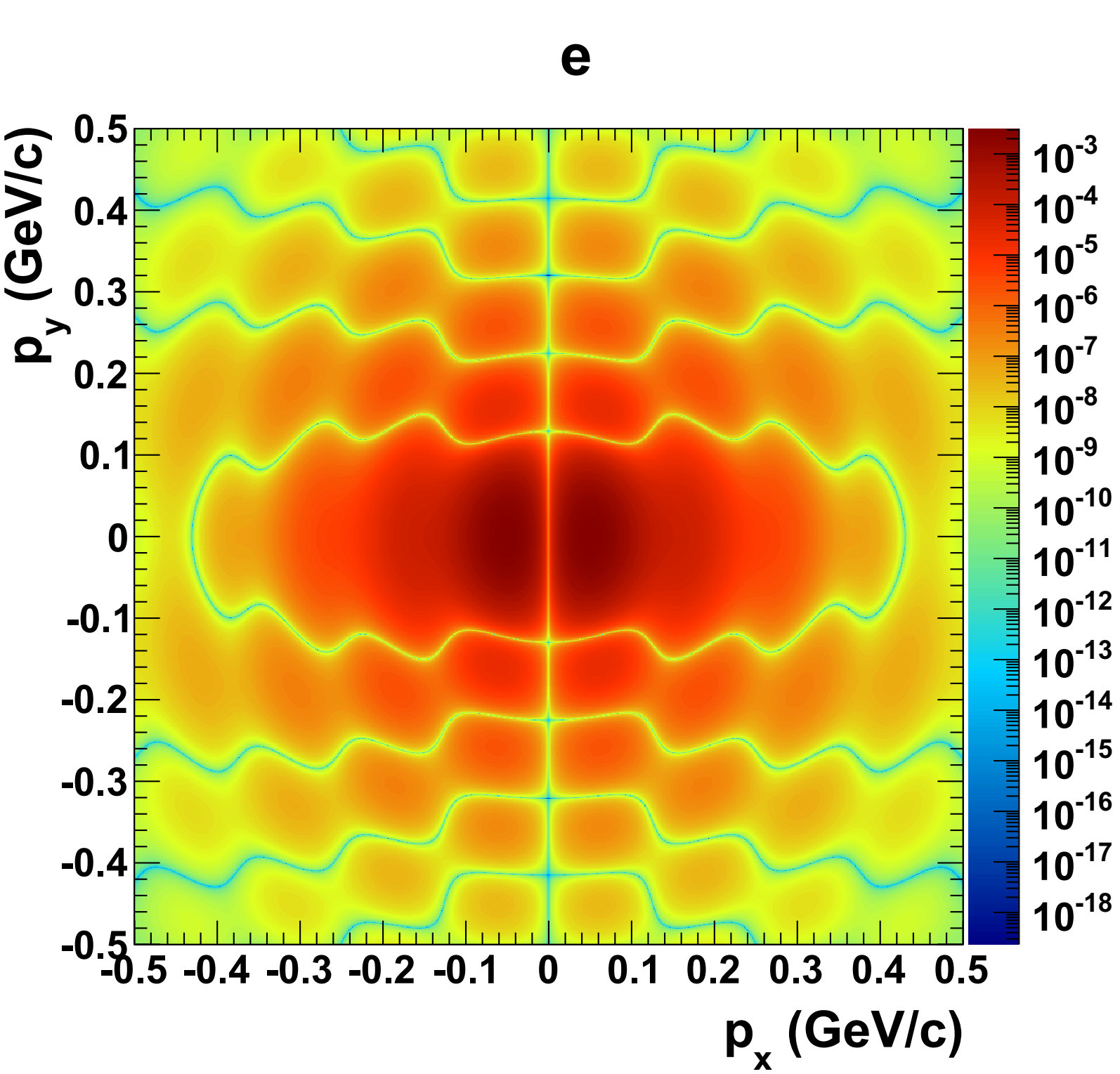}
\includegraphics[keepaspectratio,width=0.225\textwidth]{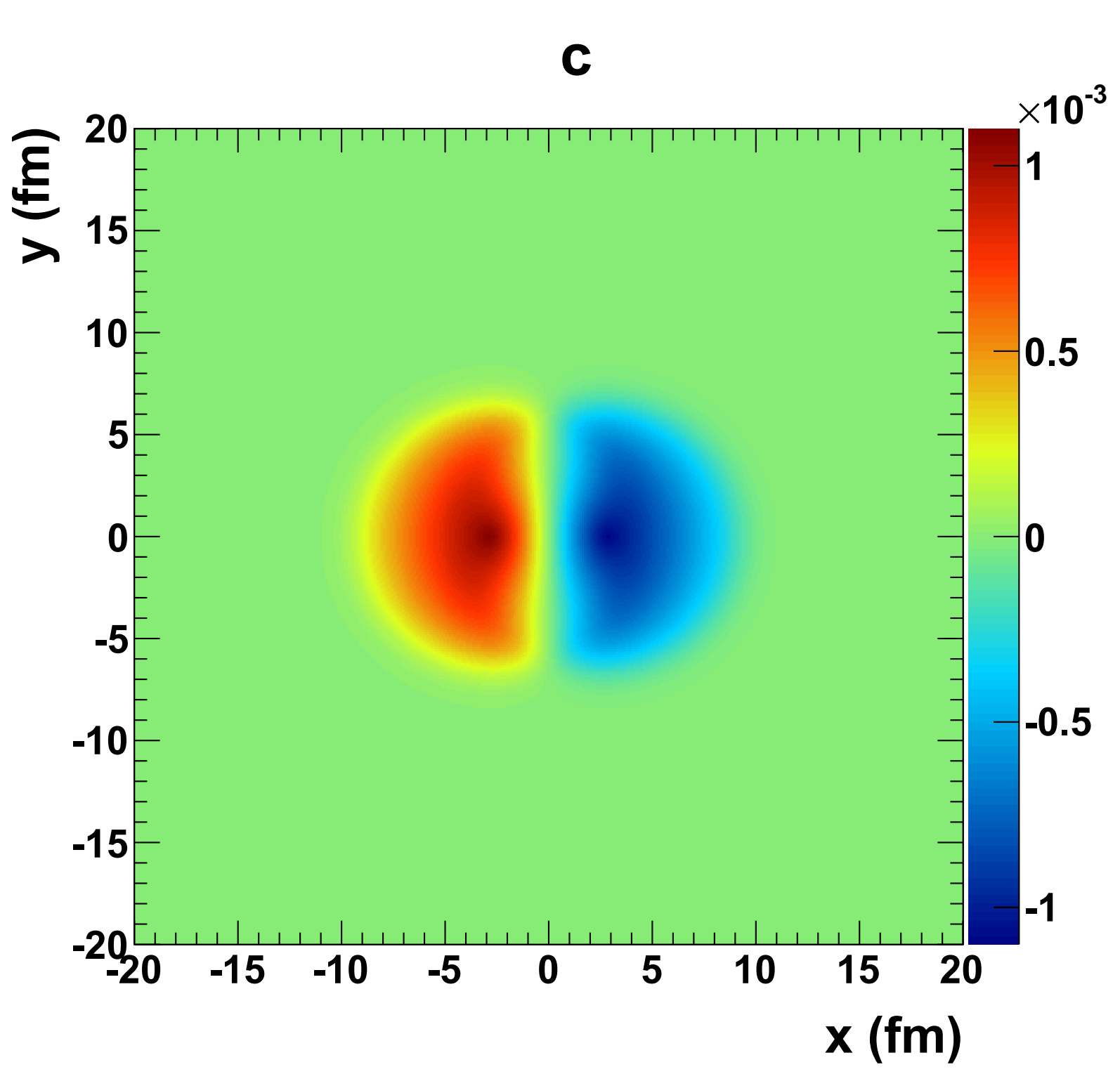}
\includegraphics[keepaspectratio,width=0.225\textwidth]{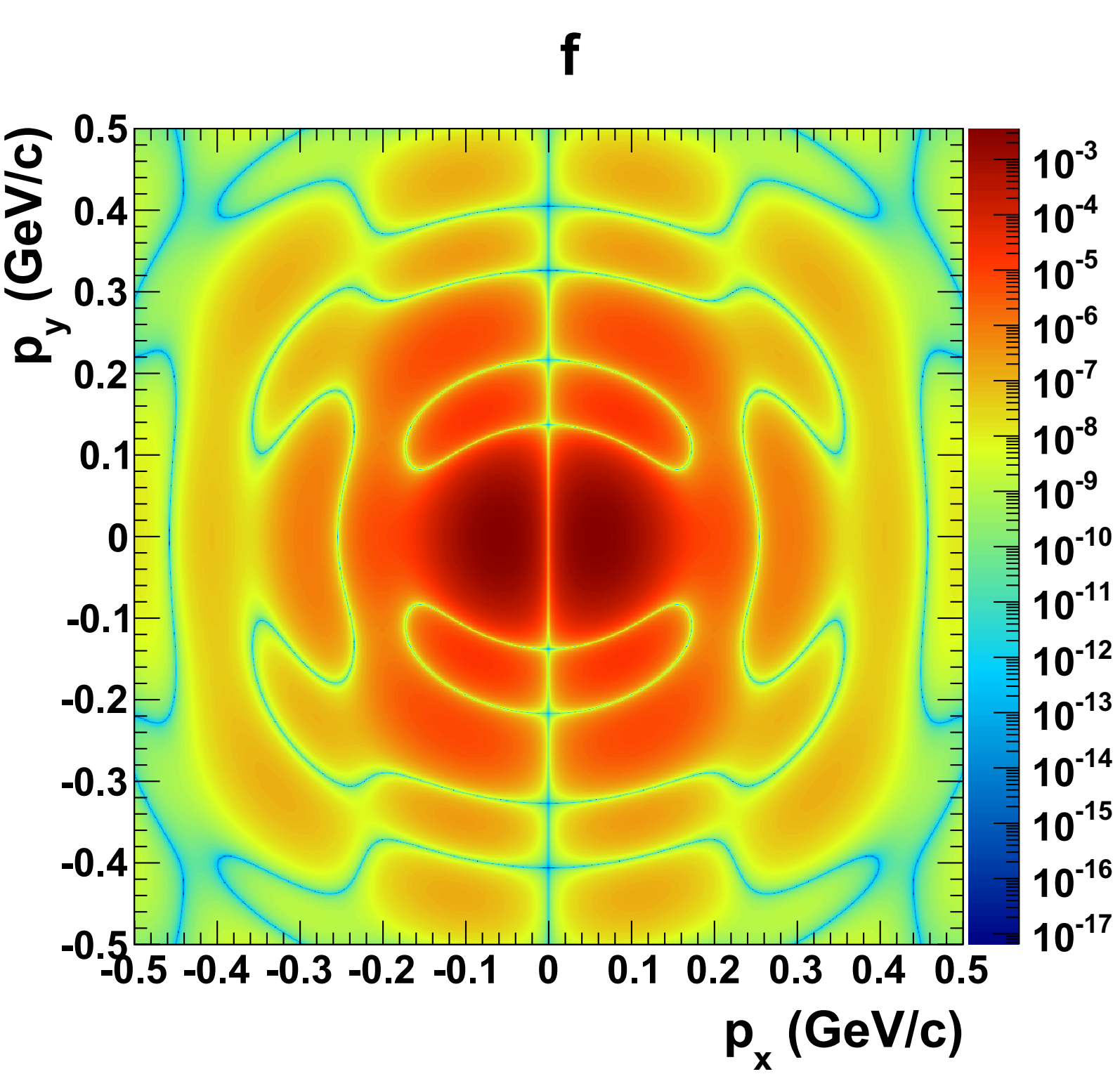}

\caption{Amplitude and momentum distribution patterns of coherent J/$\psi$ photoproduction in Au+Au collisions at $\sqrt{s_{\rm{NN}}} =$ 200 GeV for mid-rapidity (y=0) with different impact parameters. Panel a and d: b = 13.0 fm, panel b and e: b = 10.0 fm, panel c and f: b = 5.7 fm. }
\label{figure2}
\end{figure}
\begin{figure}[htbp]
\includegraphics[keepaspectratio,width=0.225\textwidth]{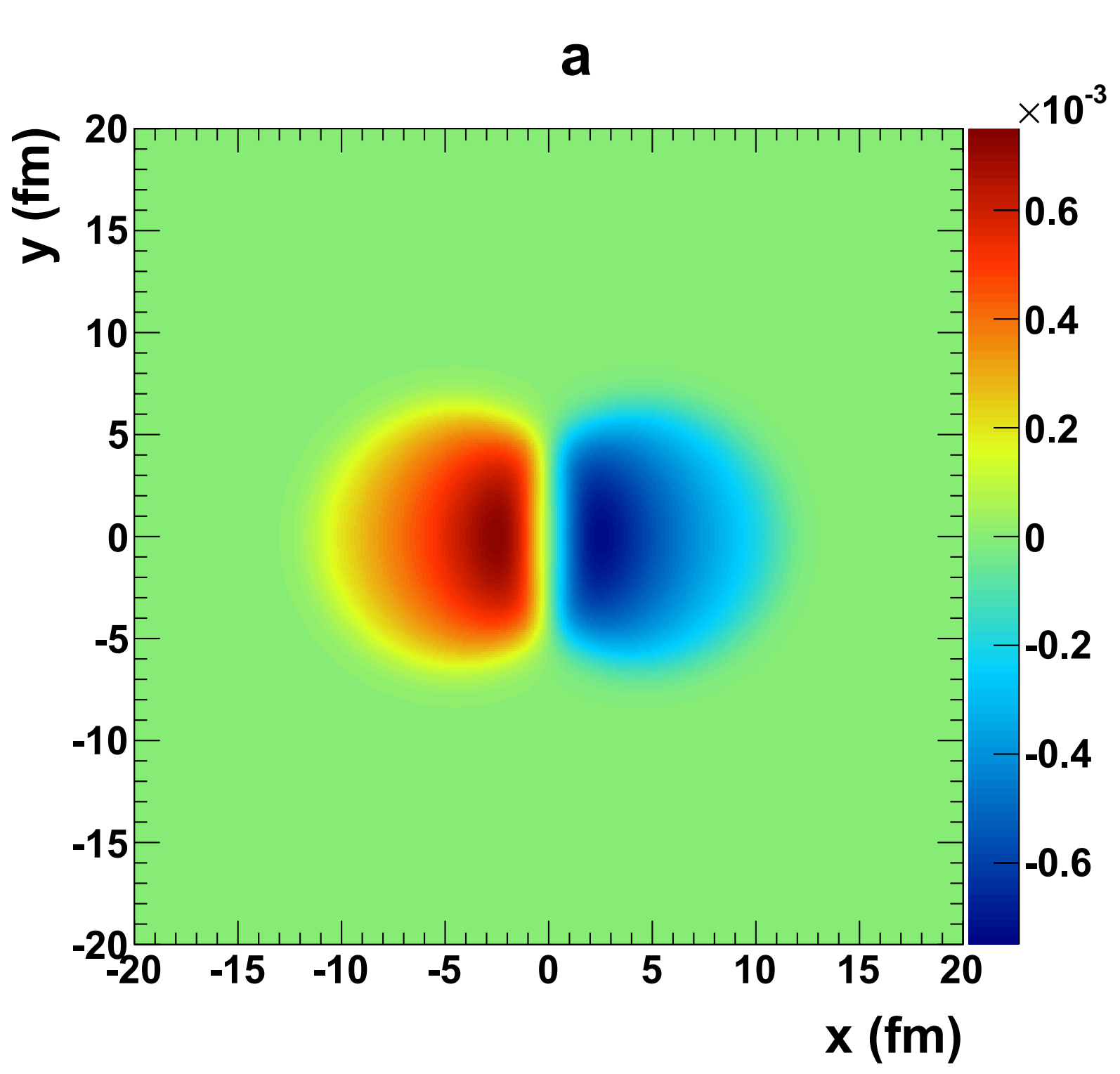}
\includegraphics[keepaspectratio,width=0.225\textwidth]{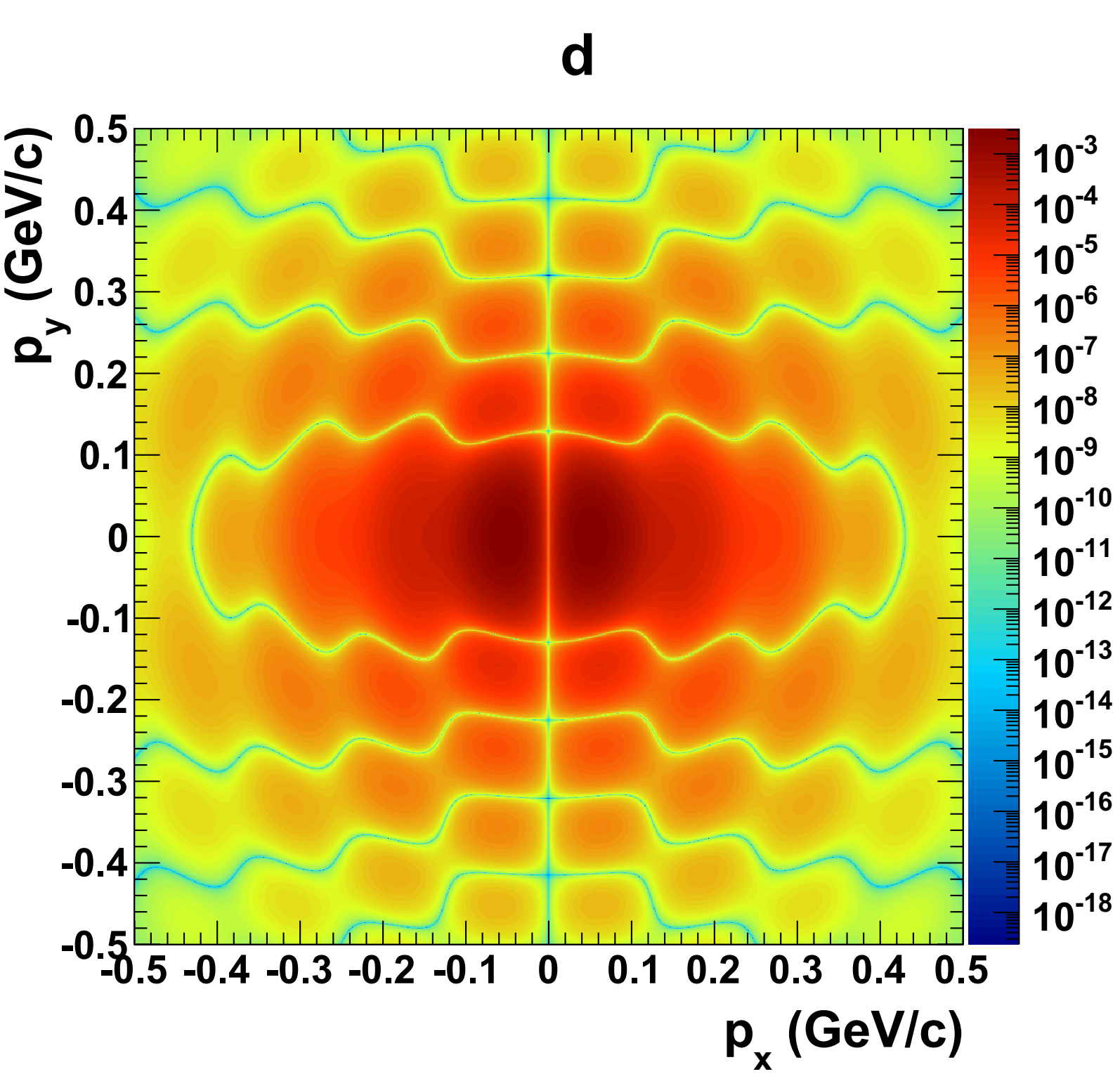}
\includegraphics[keepaspectratio,width=0.225\textwidth]{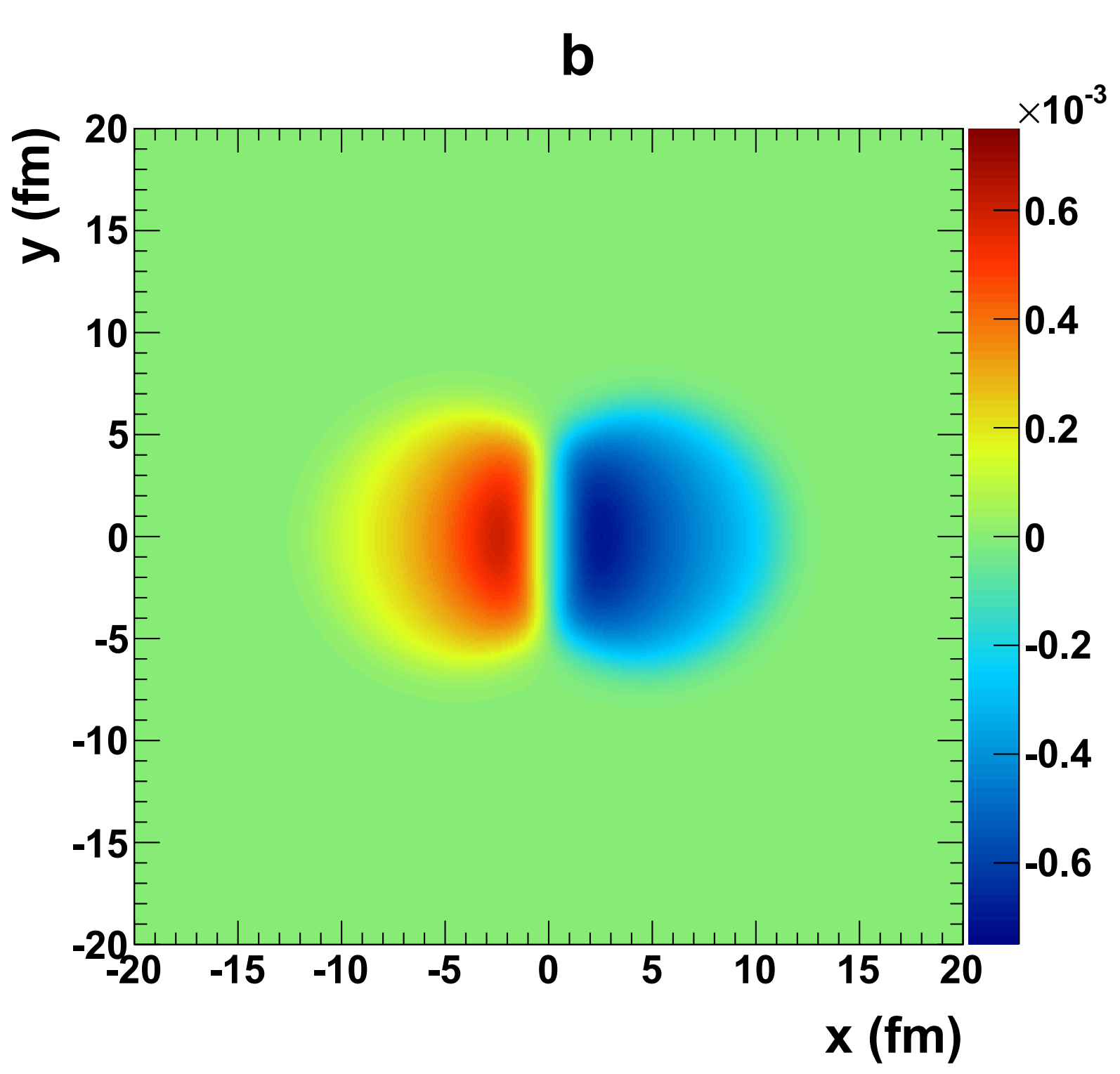}
\includegraphics[keepaspectratio,width=0.225\textwidth]{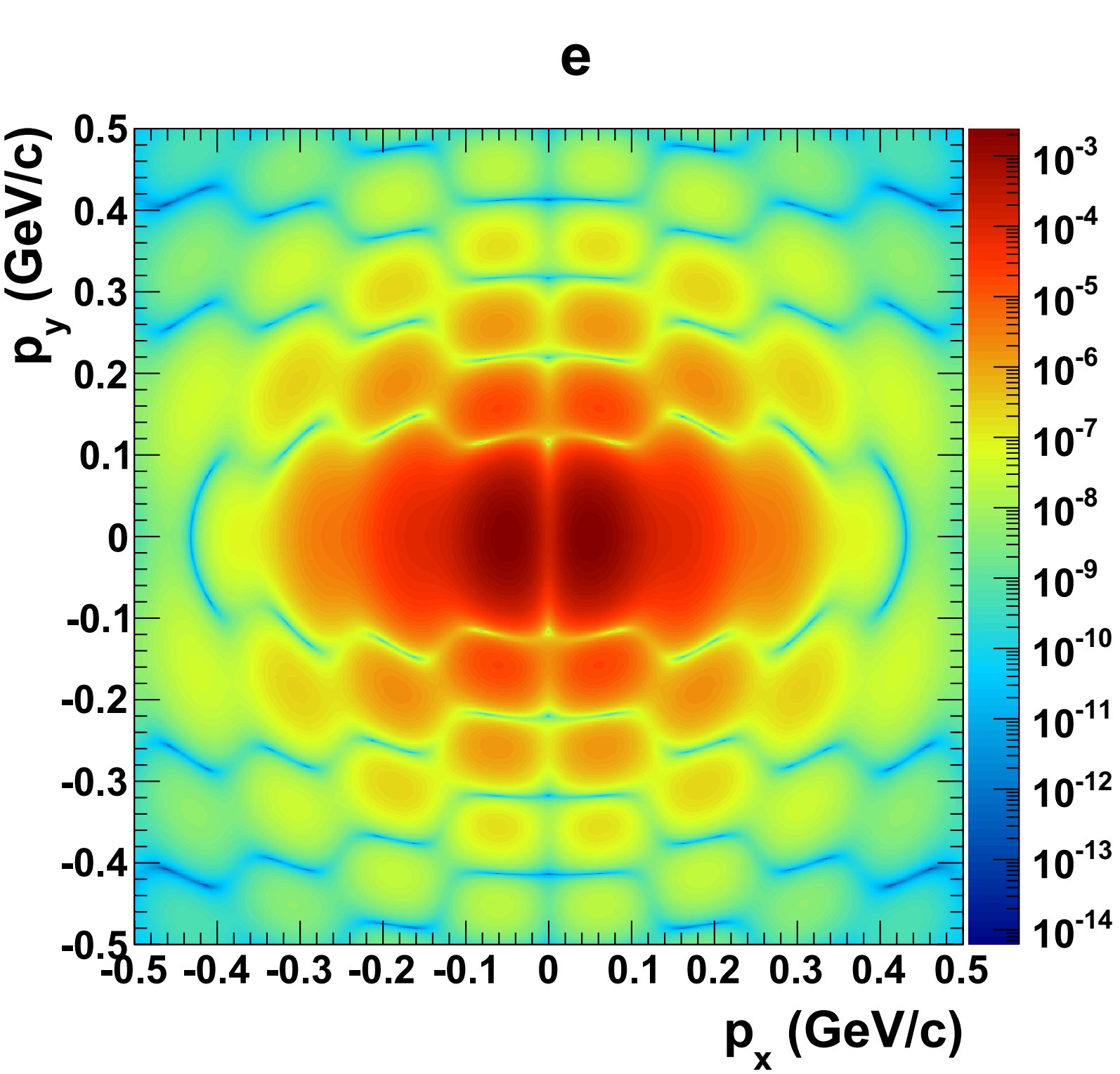}
\includegraphics[keepaspectratio,width=0.225\textwidth]{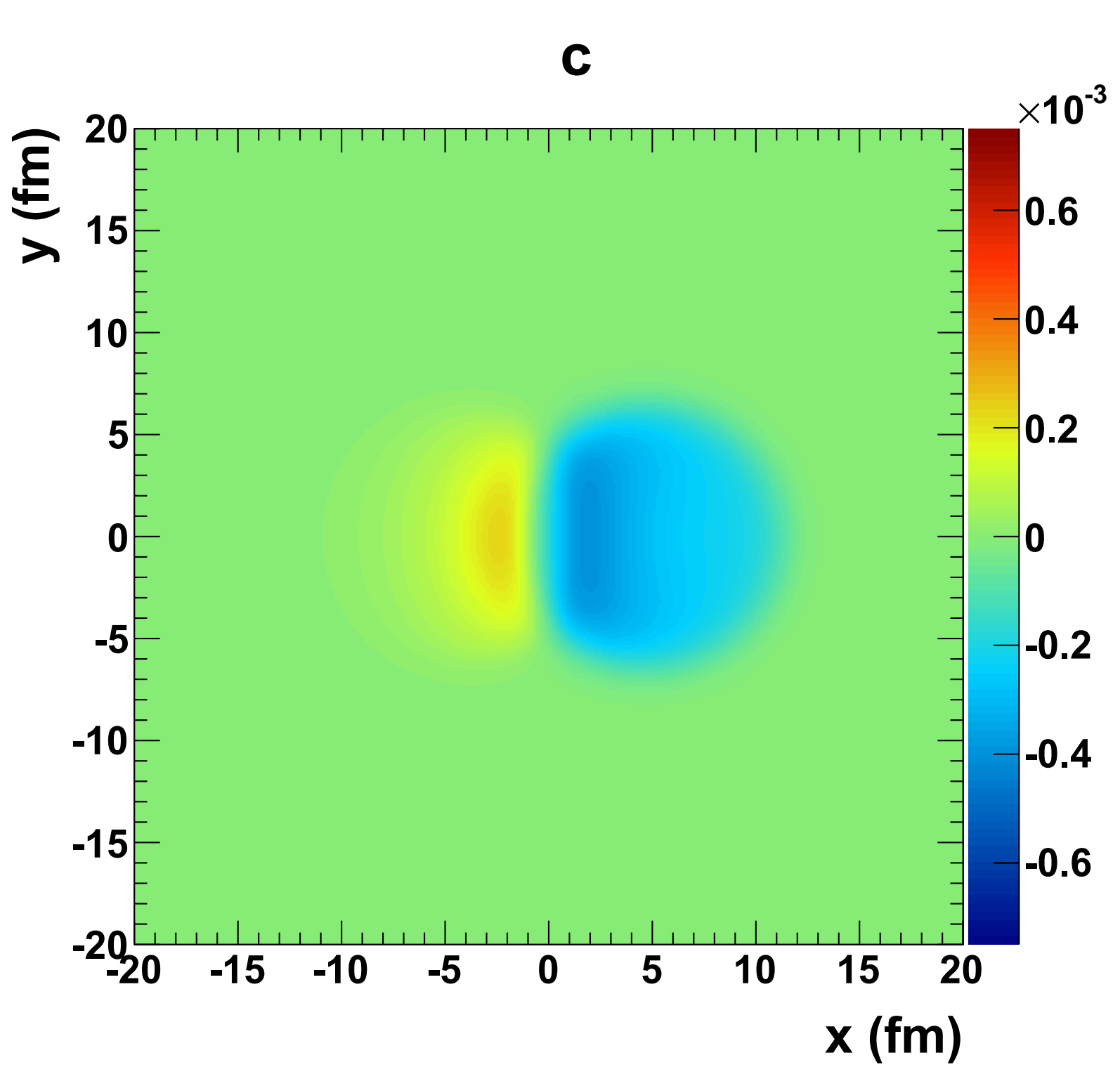}
\includegraphics[keepaspectratio,width=0.225\textwidth]{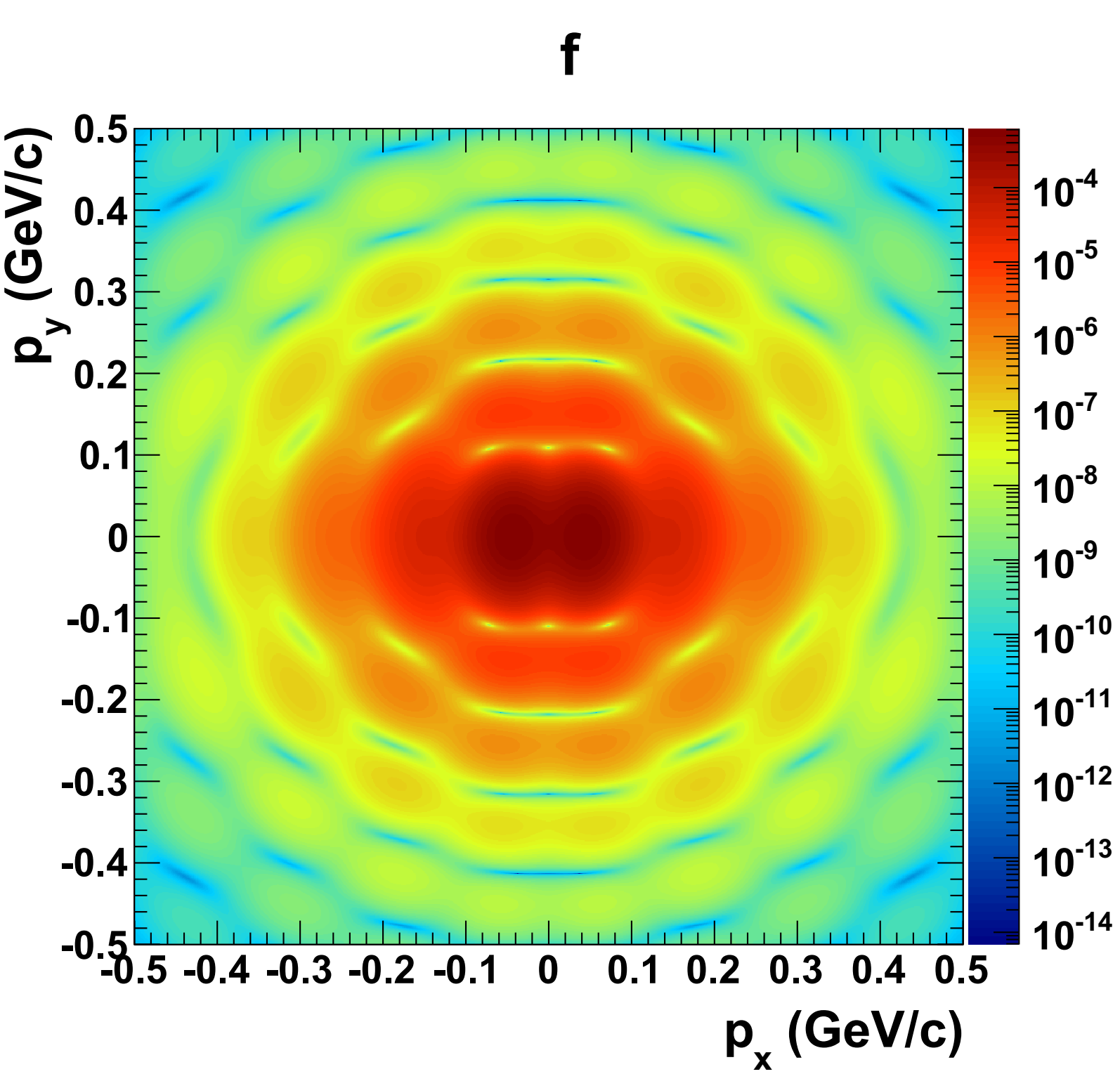}
\caption{Amplitude and momentum distribution patterns of coherent J/$\psi$ photoproduction  in Au+Au collisions at $\sqrt{s_{\rm{NN}}} =$ 200 GeV with  b = 10.0 fm at different rapidities. Panel a and d: y = 0.0, panel b and e: y = 0.6, panel c and f: y = 1.4.}
\label{figure3}
\end{figure}
\begin{figure}[htbp]
\includegraphics[keepaspectratio,width=0.225\textwidth]{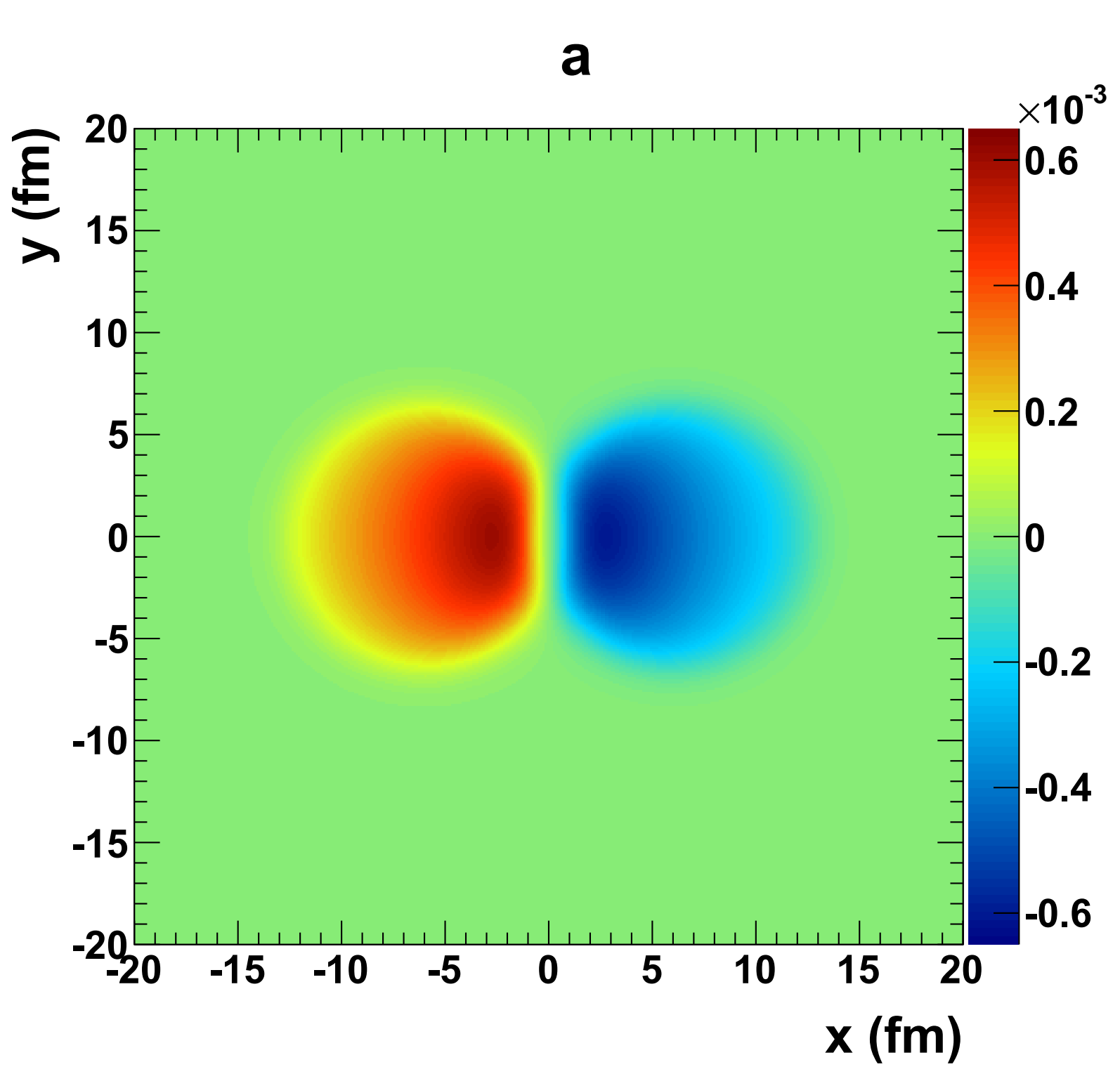}
\includegraphics[keepaspectratio,width=0.225\textwidth]{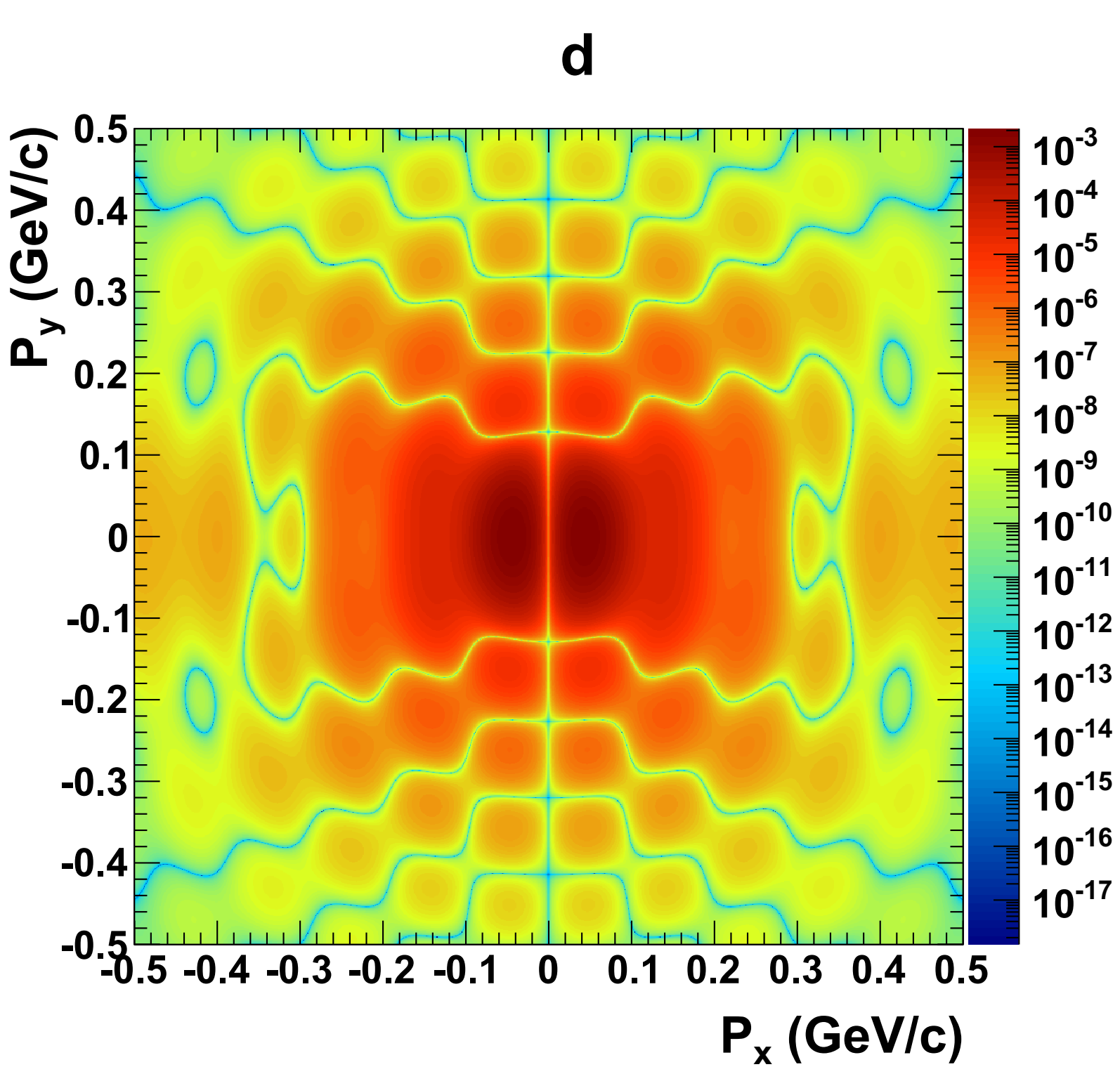}
\includegraphics[keepaspectratio,width=0.225\textwidth]{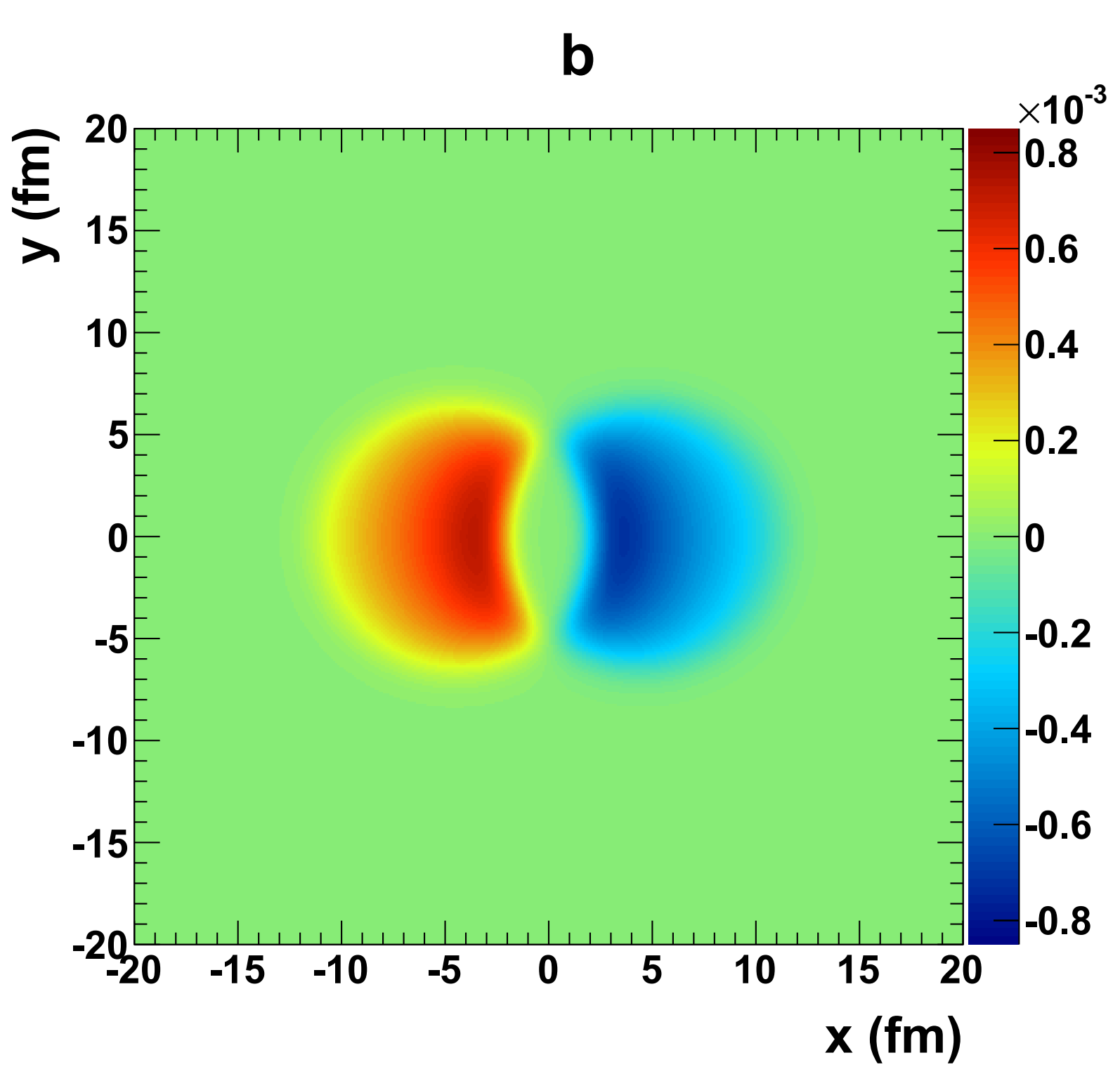}
\includegraphics[keepaspectratio,width=0.225\textwidth]{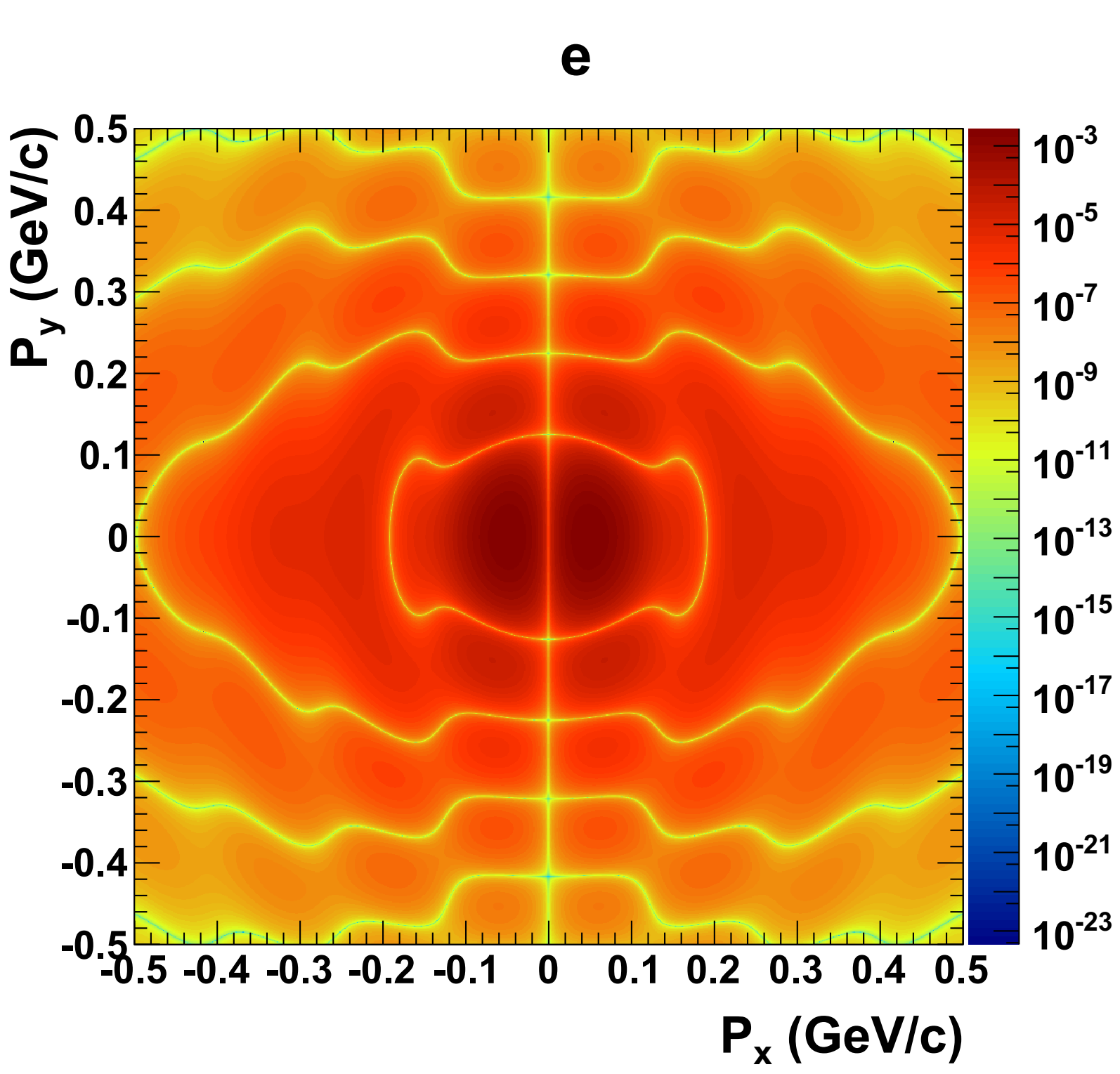}
\includegraphics[keepaspectratio,width=0.225\textwidth]{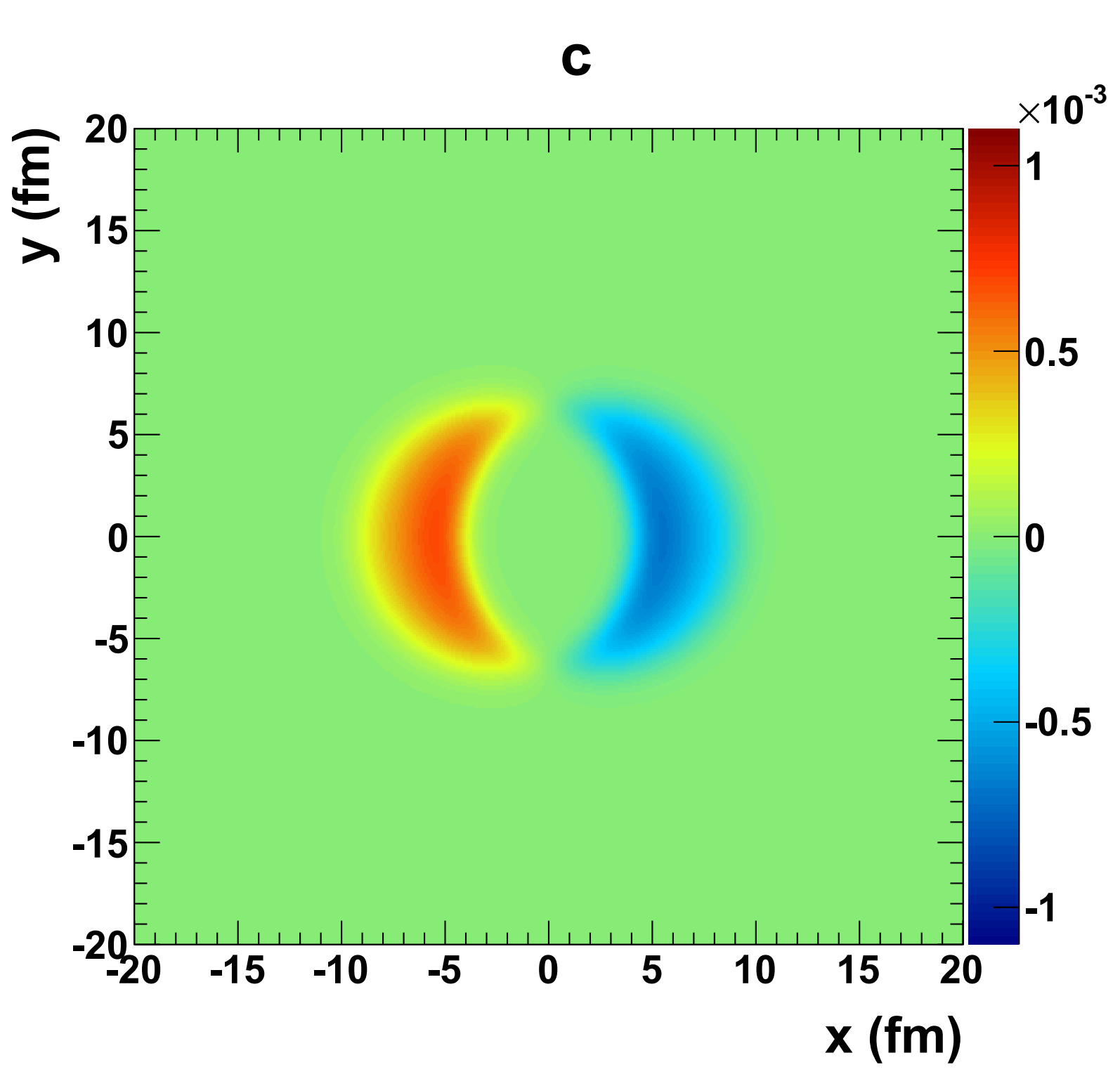}
\includegraphics[keepaspectratio,width=0.225\textwidth]{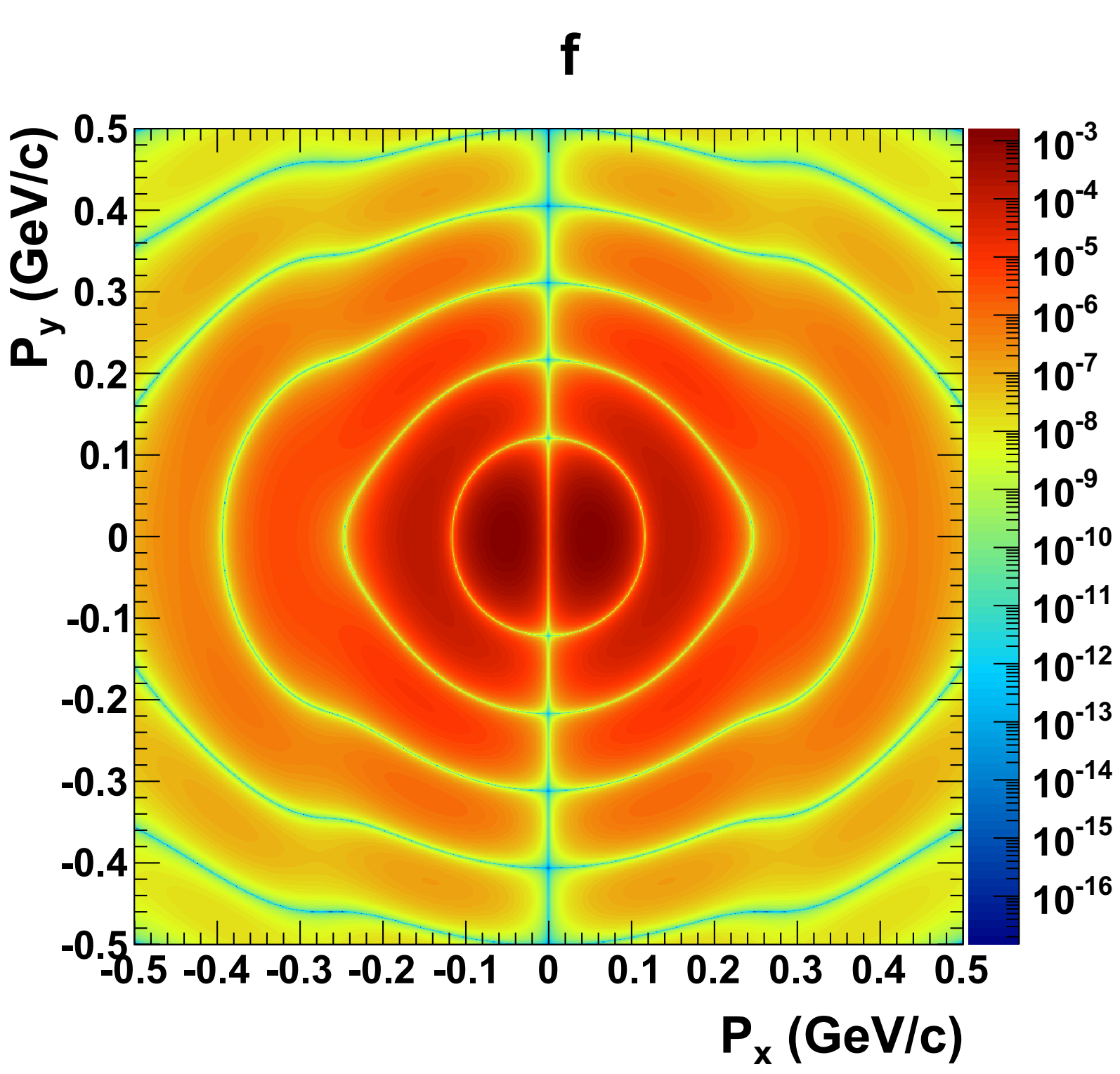}
\caption{Amplitude and momentum distribution patterns of coherent J/$\psi$ photoproduction  at mid-rapidity (y=0) in Au+Au collisions at $\sqrt{s_{\rm{NN}}} =$ 200 GeV with disruption effect from overlap region for different impact parameters. Panel a and d: b = 13.0 fm, panel b and e: b = 10.0 fm, panel c and f: b = 5.7 fm. }
\label{figure4}
\end{figure}

\begin{figure*}[htbp]
\includegraphics[keepaspectratio,width=0.9\textwidth]{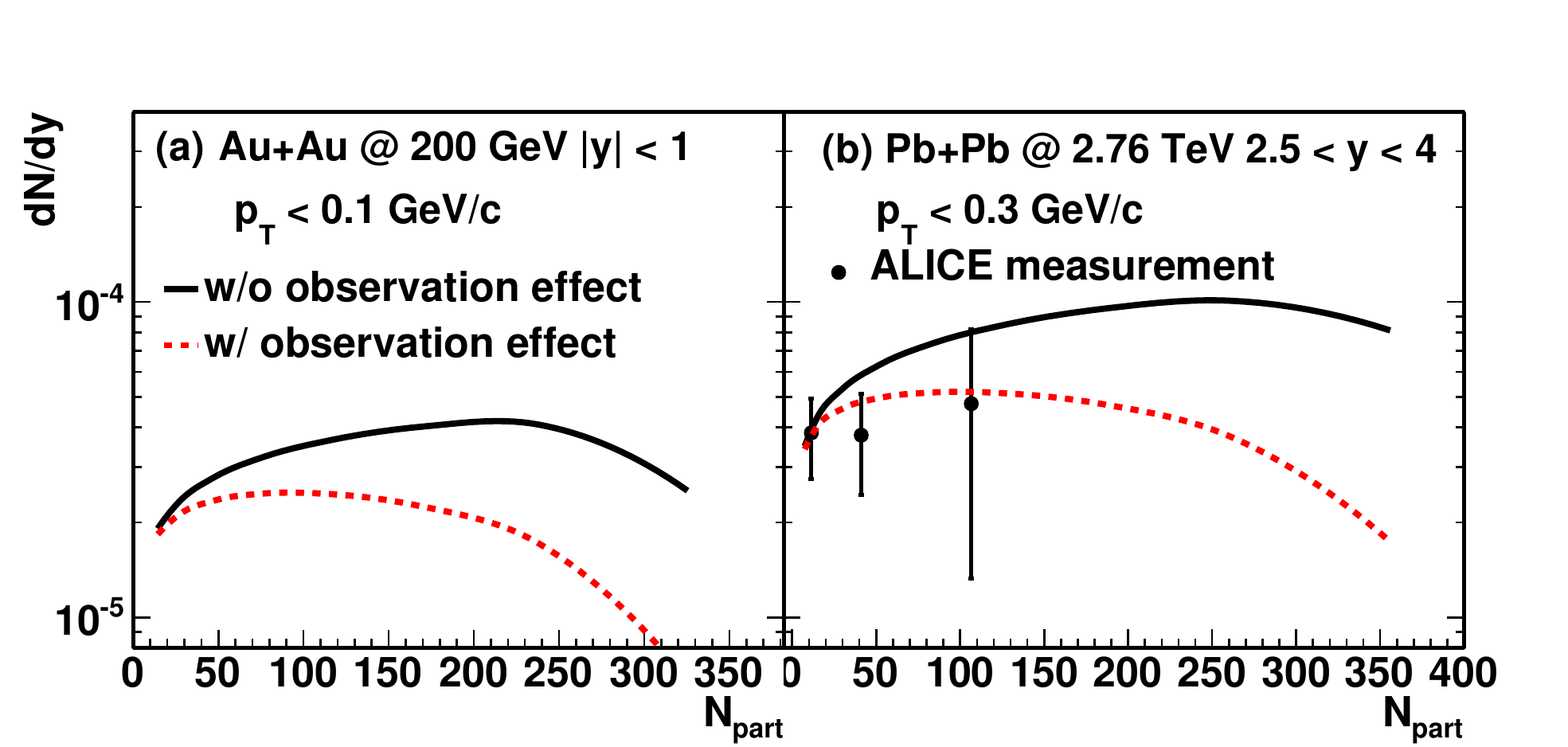}
\caption{Integrated yields of coherent J/$\psi$ production with and without ``observation'' effect as a function of $N_{\text{part}}$ in Au+Au collisions at $\sqrt{s_{\text{NN}}}$ = 200 GeV (a) and Pb+Pb collisions at $\sqrt{s_{\text{NN}}}$ = 2.76 TeV (b). Data from the ALICE experiment~\cite{LOW_ALICE} are shown for comparison. }
\label{figure5}
\end{figure*}
In heavy-ion collisions, the momentum of J$/\psi$ can be reconstructed by the detectors, which allows us to observe the interference fringes in momentum space. The probability distribution of J$/\psi$ in momentum space can be calculated by performing a Fourier transformation to the amplitude in coordinate space, defined as:
  \begin{equation}
  \frac{d^{2}P}{dp_{x}dp_{y}} = \frac{1}{2\pi} |\int d^{2}r (A_{1}(y,\textbf{r}, -\frac{b}{2}) +A_{2}(y,\textbf{r}, \frac{b}{2}))e^{i \textbf{p} \cdot \textbf{r}}|^{2},
  \label{equation1}
  \end{equation}
  where $A_{1}(y,\textbf{r}, -\frac{b}{2})$ and $A_{2}(y,\textbf{r},\frac{b}{2})$ are the amplitudes with rapidity $y$ for J$/\psi$ production at distant $\textbf{r}$ from the two colliding nuclei, respectively; b is the slit separation, which is the so-called impact parameter in heavy-ion collisions. We take $\hbar = c =1$ here. The J$/\psi$ production is coherent over the entire nucleus, for a start, we approximate the production regions as two point sources, one at the center of each nucleus. At mid-rapidity ($y = 0$), the magnitudes of amplitudes from the two directions are the same. With the point source assumption, the amplitudes from nucleus 1 and 2 at mid-rapidity can be written as:
  \begin{equation}
  A_{1}(0,\textbf{r}, -\frac{b}{2}) = \delta(r -  \frac{b}{2}),  A_{2}(0,r, \frac{b}{2}) = -\delta(r + \frac{b}{2}).
  \label{equation2}
  \end{equation}
  Here, $\delta(x)$ is the Dirac delta function. The amplitude distribution in coordinate space and the corresponding interference pattern in momentum space with slit separation $b =$ 15 fm  are shown in Fig.~\ref{figure1} panel a and d, respectively. The momentum distribution reveals typical Young's double-slit interference pattern with a series of alternating light and dark fringes. The width of the fringes reflects the distant between the two slits: $d = \frac{2\pi}{b}$. The straight fringes exhibit the central minimum due to the opposite sign amplitudes from the two sources. To more realistic case, the two nuclei (slits) possess certain size and profile at fermi scale. The density profile of nucleus can be parameterized by the Woods-Saxon distribution:
  \begin{equation}
  \rho_{A}(r)=\frac{\rho^{0}}{1+\exp[(r-R_{\rm{WS}})/d]}
  \label{equation2_new}
  \end{equation}
  where the radius $R_{\rm{WS}}$ and skin depth $d$ are based on fits to electron scattering data~\cite{0031-9112-29-7-028}, and $\rho^{0}$ is the normalization factor. With $A(\textbf{r}) =T (\textbf{r}) =\int^{+\infty}_{-\infty}\rho(\sqrt{\textbf{r}^{2} + z^{2}})dz$, the amplitude distribution in coordinate space and the corresponding momentum distribution for Au+Au collisions are shown in panel b and e, respectively. Due to the profile of slits, typical diffraction rings emerge on top of the interference fringes in momentum distribution. The amplitudes in coordinate space are further modified by the spatial photon flux, nuclear shadowing and coherent length effects. The spatial photon flux induced by ions can be given by the equivalent photon approximation~\cite{KRAUSS1997503}:
    \begin{equation}
  \label{equation2}
  \begin{aligned}
  & \frac{d^{3}N_{\gamma}(\omega_{\gamma},\vec{x}_{\bot})}{d\omega_{\gamma} d\vec{x}_{\bot}} = \frac{4Z^{2}\alpha}{\omega_{\gamma}}\bigg|\int\frac{d^{2}\vec{k}_{\gamma\bot}}{(2\pi)^{2}}\vec{k}_{\gamma\bot}\frac{F_{\gamma}(\vec{k}_{\gamma})}{|\vec{k}_{\gamma}|^{2}}
  e^{i\vec{x}_{\bot}\cdot\vec{k}_{\gamma\bot}}\bigg|^{2}
  \\
  & \vec{k}_{\gamma}=(\vec{k}_{\gamma\bot},\frac{\omega_{\gamma}}{\gamma_{c}}) \ \ \ \ \  \omega_{\gamma}=\frac{1}{2}M_{\text{J}/\psi} e^{\pm y}
  \end{aligned}
  \end{equation}
  where $\vec{x}_{\bot}$ and $\vec{k}_{\gamma\bot}$ are 2-dimensional photon position and momentum vectors perpendicular to the beam direction, $Z$ the nuclear charge, $\alpha$ the electromagnetic coupling constant, $\gamma_{c}$ the Lorentz factor of the photon-emitting nucleus, $M_{\text{J}/\psi}$ and $y$ the mass and rapidity of J/$\psi$, and $F_{\gamma}(\vec{k}_{\gamma})$ the nuclear electromagnetic form factor.  $F_{\gamma}(\vec{k}_{\gamma})$ is obtained via the Fourier transformation of the charge density in the nucleus. The calculation of scattering amplitude $\Gamma_{\gamma A \rightarrow \rm{J}/\psi A}(\textbf{r})$ with shadowing effect can be performed with quantum Glauber~\cite{doi:10.1146/annurev.nucl.57.090506.123020} + vector meson dominance (VMD) approach~\cite{RevModPhys.50.261} coupled with the parameterized forward scattering amplitude $f_{\gamma p \rightarrow \rm{J}/\psi N}(0)$~\cite{Klein:2016yzr} as input:
  \begin{equation}
\Gamma_{\gamma A \rightarrow \rm{J}/\psi A}(\textbf{r}) = \frac{f_{\gamma p \rightarrow \rm{J}/\psi N}(0)}{\sigma_{\rm{J}/\psi N}} \times 2 \times [1-\rm{exp}(-\frac{\sigma_{\rm{J}/\psi N}}{2} \times T^{\prime}(\textbf{r})].
  \label{equation4}
  \end{equation}
Using the optical theorem and VMD relation, the total cross section for J$/\psi N$ scattering is given by:
\begin{equation}
\sigma_{\rm{J}/\psi N} = \frac{f_{\rm{J}/\psi}}{4\sqrt{\alpha}C}f_{\gamma p \rightarrow \rm{J}/\psi N}(0),
  \label{equation5}
\end{equation}
where $f_{\rm{J}/\psi}$ is J$/\psi$-photon coupling and C is a correction factor for the non-diagonal coupling through higher mass vector
mesons~\cite{HUFNER1998154}. To account for the coherent length effect, $T^{\prime}(\textbf{r})$ is defined as:
\begin{equation}
 T^{\prime}(\textbf{r}) = \int^{+\infty}_{-\infty} dz \rho(\sqrt{\textbf{r}^{2} + z^{2}}) e^{iq_{L}z}, q_{L} = \frac{M_{\rm{J}/\psi} e^{y}}{2\gamma}
  \label{equation6}
\end{equation}
where $q_{L}$ is the longitudinal momentum transfer required to produce a real J$/\psi$. With these effects, the amplitude from one direction can be written as:
\begin{equation}
 A_{1}(y,\textbf{r}, -\frac{b}{2}) = \Gamma_{\gamma A \rightarrow \rm{J}/\psi A}(\textbf{r}-\frac{\textbf{b}}{2}) \times \sqrt{\frac{d^{2}N_{\gamma}(\textbf{r}-\frac{\textbf{b}}{2})}{d^{2}r}},
  \label{equation7}
\end{equation}
which makes the slits become asymmetric. The amplitude distributions of the asymmetric slits in Au+Au collisions at $\sqrt{s_{\rm{NN}}} =$ 200 GeV for mid-rapidity (y=0) are shown in Fig.~\ref{figure1} panel c and the resulting interference pattern is depicted in panel f. The corresponding diffraction pattern does not show typical symmteric diffraction rings due to the asymmetric profile of slits, and the interference fringes become curving.

The coherent photoproduction has been studied in detail in ultra-peripheral collisions (UPCs)~\cite{UPCreview}, in which the impact parameter is larger than twice the nuclear radius. The destructive interference of $\rho^{0}$ in transverse momentum for UPCs has been proposed by Klein and Nystrand~\cite{UPC_PT}, and verified by the STAR collaboration~\cite{Abelev:2008ew}. However, no special azimuthal direction can be determined in UPCs, which prevents us to observed the two-dimensional interference fringes. Fortunately, in hadronic heavy-ion collisions (HHICs), when the two nuclei overlap, the reaction plane, spanned by the impact parameter and the beam axis,  can be estimated from the azimuthal anisotropy of produced particles due to the asymmetric geometry in overlap region. Herein, the disruption to interference action from the violent strong interactions in overlap region is ignored, which would be detailedly discussed in the later. Furthermore, in HHICs, the impact parameter (centralities) could be precisely determined, which allows us to vary the slit separation in the presented scenario. The profile of amplitudes and the corresponding momentum distributions in Au+Au collisions at $\sqrt{s_{\rm{NN}}} =$ 200 GeV with different slit separations at mid-rapidity (y=0) are shown in Fig.~\ref{figure2}. As demonstrated in the figure, the band width of interference fringes becomes wider when the slit separation gets smaller. And the diffraction pattern also changes due to the different profile of slits with different slit separations. In the proposed scenario, the relative brightness between the two slits could also be adjusted by detecting the coherently produced J$/\psi$'s at different rapidities. Figure~\ref{figure3} shows the amplitude distributions and the corresponding momentum distributions in Au+Au collisions at $\sqrt{s_{\rm{NN}}} =$ 200 GeV with the slit separation $b = 10$ fm for J$/\psi$ production at different rapidities. When going to forward rapidities, due to the dominant production from one direction, the interference fringes merge into each other, which makes the diffraction pattern dominant over the momentum distributions.

A well-known double-slit thought experiment predicts that if particle detectors are positioned at the slits to solve the ``which way'' problem, the interference pattern will disappear, which illustrates the complementary principle that matter can behave as either particle or wave, but can not possess both at the same time. Conventional wisdom says that, in HHICs, the strong interactions and possible quark-gluon-plasma (QGP)~\cite{PBM_QGP} in overlap region detect the coherent produced J$/\psi$'s and determine the ``which way'' information, leading to decoherence, which prohibits the coherent photoproduction. However, ALICE~\cite{LOW_ALICE} and STAR~\cite{1742-6596-779-1-012039} collaborations observed significant anomaly excess of J$/\psi$ production in HHICs at very low transverse momentum ($p_{T} < $ 0.3 GeV/c), which could be qualitatively described by coherent photoproduction mechanism~\cite{PhysRevC.97.044910}. The strong evidence of existence of coherent photoproduction in HHICs focus us to consider the disruption from overlap region more carefully. The strong interactions are short-range, so they can not observe or detect the coherent photoproduction out of the hot medium region. The ``which way'' information is only obtained within the extent of strong interactions, which means that the ``which slit'' problem is only partially solved or the two slits are only sectionally blocked by the interactions but not all. Herein, we adopt a simple assumption that the slits are only blocked in the initial overlap region. The amplitude distributions of the sectional blocked slits and the corresponding momentum distribution patterns in Au+Au collisions at $\sqrt{s_{\rm{NN}}} =$ 200 GeV with different impact parameters at mid-rapidity are depicted in Fig.~\ref{figure4}. In comparison with the interference and diffraction patterns shown in Fig.~\ref{figure2}, the diffraction rings and interference fringes change significantly due to the partially blocking of the slits. The integrated yields of coherent J/$\psi$ production as a function of $N_{\rm{part}}$ with the whole slits and sectional blocked slits (``observation'' effect) are shown in Fig.~\ref{figure5} for Au+Au collisions at $\sqrt{s_{\text{NN}}}$ = 200 GeV (a) and Pb+Pb collisions at $\sqrt{s_{\text{NN}}}$ = 2.76 TeV (b). The integrated production rate is significantly reduced by the ``observation'' effect in semi-central and central collisions. And in comparison with the experimental measurements from ALICE~\cite{LOW_ALICE} and STAR~\cite{1742-6596-779-1-012039}, the data seem to favor the calculation with ``observation'' effect , however could not distinguish the two scenarios due to limited statistics.  

By making use of the coherent photoproduction in HHICs, one can take Young's famous experiment one step further at fermi scale and create a truly one-by-one double-slit set-up comprised of new entities -- vector meson. The slit separations and relative brightness between slits can be adjusted in presented scenario. In addition, the strong interactions in the overlap region detect the coherent products and partially solve the ``which-way'' information, which serves as a good double-slit scenario to test the ``observation'' effect on wave-particle duality and further demonstrates the complementary principle.

We thank Prof. Spencer Klein and Pengfei Zhuang for useful discussions. This work was funded by the National Natural Science Foundation of China under Grant Nos. 11775213, 11505180, and 11375172, the U.S. DOE Office of Science under contract No. DE-SC0012704, and MOST under Grant No. 2014CB845400 and 2016YFE0104800.

\nocite{*}
\bibliographystyle{aipnum4-1}
\bibliography{aps}
\end{document}